\documentclass{article}
\usepackage{arxiv}
\usepackage{amsmath}
\usepackage{amssymb}
\usepackage[utf8]{inputenc} 
\usepackage[T1]{fontenc}    
\usepackage{amsfonts}       
\usepackage{graphicx}
\usepackage{stackrel}
\usepackage{cite}

\usepackage{authblk}
\title{Bayesian inference of L\'evy walks via hidden Markov models}
\author[1]{Seongyu Park \raggedright}
\author[2, 3]{Samudrajit Thapa}
\author[1]{Yeongjin Kim}
\author[4, \ddag]{Michael A. Lomholt}
\author[1, \S]{Jae-Hyung Jeon}
\affil[1]{%
Department of Physics, Pohang University of Science and Technology (POSTECH), Pohang 37673, Republic of Korea}
\affil[2]{%
Sackler Center for Computational Molecular and Materials Science, Tel Aviv University, Tel Aviv 69978, Israel}
\affil[3]{%
School of Mechanical Engineering, Tel Aviv University, Tel Aviv 69978, Israel}
\affil[4]{%
PhyLife, Department of Physics, Chemistry and Pharmacy,
University of Southern Denmark, Campusvej 55, 5230 Odense M, Denmark}
\affil[$\ddag$]{%
E-mail: mlomholt@sdu.dk}
\affil[$\S$]{%
E-mail: jeonjh@postech.ac.kr}
\date{}                     

\begin{document}
\maketitle
\begin{abstract}
\textbf{Abstract.}  The L\'evy walk is a non-Brownian random walk model that has been found to describe anomalous dynamic phenomena in diverse fields ranging from biology over quantum physics to ecology. Recurrently occurring problems are to examine whether observed data are successfully quantified by a model classified as L\'evy walks or not and extract the best model parameters in accordance with the data. Motivated by such needs, we propose a hidden Markov model for L\'evy walks and computationally realize and test the corresponding Bayesian inference method.
We introduce a Markovian decomposition scheme to approximate a renewal process governed by a power-law waiting time distribution. Using this, we construct the likelihood function of L\'evy walks based on a hidden Markov model and the forward algorithm. With the L\'evy walk trajectories simulated at various conditions, we perform the Bayesian inference for parameter estimation and model classification. We show that the power-law exponent of the flight-time distribution can be successfully extracted even at the condition that the mean-squared displacement does not display the expected scaling exponent due to the noise or insufficient trajectory length. It is also demonstrated that the Bayesian method performs remarkably inferring the L\'evy walk trajectories from given unclassified trajectory data set if the noise level is moderate.
\end{abstract}

\section{Introduction}
Complex diffusion dynamics is often observed in diverse fields, such as transport dynamics of living macromolecules inside a biological cell~\cite{metzler2014, saxton1997, golding2006, jeon2011, rienzo2014, krapf2019, lee2021, park2021}, tracer particles in polymer networks~\cite{wang2012, cherstvy2019, kim2020}, foraging dynamics of living organisms~\cite{ariel2015, ipina2019, patteson2015} or animals~\cite{humphries2010, jager2011}, and stock prices in the stock market~\cite{wergen2011, plerou2000}, to name a few. Such complex dynamics usually deviate from Brownian dynamics~\cite{brown1828, einstein1956, smoluchowski1906} characterized by Gaussian displacements $P(\Delta x; \Delta t) \propto \exp \left (-\frac{\Delta x^2}{4D\Delta t}\right )$ and the linear growth of the mean-squared displacement~(MSD) with time $\langle x^2(t) \rangle \propto t$. They exhibit non-Brownian characteristics in which the MSD usually increases with time as
\begin{equation} \label{eq:msd}
    \langle x^2 (t) \rangle \propto t^\alpha
\end{equation}
with the anomaly exponent $\alpha$ in the range $0<\alpha \leq 2$~\cite{reverey2015, lampo2017} and/or the displacements have a non-Gaussian distribution that violates the central limit theorem~\cite{wang2012, lampo2017}. These anomalous diffusion dynamics have been explained by continuous-time random walk~\cite{scher1975, montroll1969},  fractional Brownian motion~\cite{mandelbrot1968}, L\'evy walk~\cite{zaburdaev2015}, annealed transient time~\cite{massignan2014}, and scaled Brownian motion~\cite{lim2002, jeon2014}.

It is often challenging to pinpoint the anomalous diffusion model underlying observed complex diffusion dynamics and determine the value of model parameters (e.g., the anomaly exponent). The former is understood as model classification and the latter as parameter estimation. To get proper information on these tasks, conventionally, one analyzes the experimental time-series data with various dynamic measures such as MSD, Van-Hove correlation functions, step-length or flight-time distribution, and  ergodicity breaking parameter~\cite{metzler2009analysis, Jeon2013, safdari2015, cherstvy2015, burov2011, ernst2014, kepten2015}.  
However, the interpretation of the data can be subjective and sometimes becomes highly nontrivial if the given data are not ideal. The non-ideal conditions include insufficient samples in terms of length and number, the high noise level in the data, and the spatiotemporal heterogeneity in the samples. To overcome these difficulties, recently, new approaches, e.g., Bayesian inference and machine learning, have been developed with keen interest from cross-disciplinary sciences  ~\cite{jamali2021, gorka2020, cichos2020, granik2019, martin2016, thapa2018, cherstvy2019, krog2018, krog2017, auger2015}.

In this work, we develop a Bayesian inference method for analyzing L\'evy walks. Bayesian inference provides a direct comparison between the possible statistical models and finds the most likely model parameters over a given parameter space by calculating the likelihood functions of given data. 
In a Bayesian framework, one does not have to define point-estimators and extract the statistical features (e.g., MSD or van-Hove correlation function). From a single raw trajectory, diffusion models and model parameters can be examined to derive the best-fit model and its model parameters.
Recently, some efforts have been devoted to developing the Bayesian framework for Brownian motion, fractional Brownian motion, and diffusing diffusivity~\cite{martin2016, thapa2018, cherstvy2019, krog2018, krog2017}. These studies have proven the success of the Bayesian inference approach in the trajectory analysis. 

The L\'evy walk model is a physical correction of L\'evy flights~\cite{zaburdaev2015}. The corresponding trajectory consists of independent ballistic flights with constant speed $v$ and the random flight times governed by a L\'evy distribution. 
L\'evy walks are known to model various non-equilibrium dynamic processes in biology, ecology, physics, and economics. Prominent examples include foraging dynamics of bacteria, animals, and humans~\cite{raichlen2014, reynolds2009, garg2021, wosniack2017, rhee2011, focardi2009, jager2011, huo2021}, active transport of macromolecules inside a living cell~\cite{song2018, gal2010, chen2015}, blinking dynamics of quantum dots~\cite{stefani2009}, light in inhomogeneous media~\cite{barthelemy2008}, a flow motion in a rotating annulus~\cite{solomon1993}, and related Hamiltonian systems~\cite{klafter1994, geisel1985}. For modeling animal's foraging dynamics in ecology, a Bayesian inference approach was applied to differentiate L\'evy flights/walks from composite correlated random walks via calculating their likelihood functions~\cite{auger2015}. Their method, however, was based on manually-extracted step lengths and turning angles from trajectories. Moreover, the correlation between the step length and flight time---the essential ingredient of a L\'evy walk---was not taken into account in the likelihood function. Previously, we developed Bayesian inference frameworks for several anomalous diffusion models~\cite{thapa2018, cherstvy2019, krog2018, krog2017}. In this work, based on these experiences, we set the Bayesian inference method for L\'evy walks. Our approach is based on a Markov decomposition method that mathematically describes a L\'evy walk process without the manually-extracted distribution features. Using this method, we construct the likelihood function of a L\'evy walk trajectory systematically as a function of trajectory length, the power-law exponent of the flight-time distribution, and the noise strength.  

The present paper is organized in the following structure. In Sec.~\ref{sec2} we give a brief overview of Bayesian inference theory and the L\'evy walk model that are used throughout the paper. In Sec.~\ref{sec:modeling}, we propose a theory of the hidden Markov model for L\'evy walks. This model is employed to compute the likelihood function of a one dimensional L\'evy walk trajectory based on the forward algorithm (we comment on a generalization to two dimensions in the \ref{sec:2dlevy}).  Secs.~\ref{sec:result} \&~\ref{sec:results2} present the computational results of our Bayesian inference method. Here, we demonstrate that random flight times governed by a power-law distribution are generated from the hidden Markov model. We also visualize the numerically calculated likelihood functions at various conditions. We then systematically investigate the performance of our Bayesian inference tool for extracting the power-law exponent of L\'evy walk trajectories upon varying the trajectory length, the exponent $\alpha$, and the signal-to-noise ratio. In Sec.~\ref{sec:results2}, we study the model classification using the Bayesian method differentiating among L\'evy walk, fractional Brownian and scaled Brownian trajectories at various conditions. Finally, in Sec.~\ref{sec:discussion}
we summarize our study and discuss the generalization of our Bayesian inference theory of L\'evy walks for the application to other similar anomalous dynamic processes.


\section{Theoretical background} \label{sec2}
\subsection{Bayesian inference method}\label{sec:bayesian}

In this subsection, we provide a short overview of Bayesian inference in connection with our Bayesian approach to L\'evy walks in the following section. Here we introduce key concepts as well as define essential probability distributions to be used later. For a complete overview of Bayesian inference, see~Ref.~\cite{gelman2013}. Suppose that a data set $\mathcal{D}$ is given, and we have several statistical models $\mathcal{M}_i$ to explain the data. For a quantitative comparison between the models, in principle, we need to estimate the conditional probability $P(\mathcal{M}_i | \mathcal{D})$ that the model $\mathcal{M}_i$ is selected to explain $\mathcal{D}$. The best-fit model is the one that maximizes the given conditional probability.
According to Bayes' theorem~\cite{Bayes1763},  $P(\mathcal{M}_i | \mathcal{D})$ is given by
\begin{equation}
    P(\mathcal{M}_i | \mathcal{D}) = \frac{P(\mathcal{D} | \mathcal{M}_i) P(\mathcal{M}_i)}{P(\mathcal{D})}
\end{equation}
where $P(\mathcal{M}_i)$ is the prior probability that the model $\mathcal{M}_i$ is given before our observation. $P(\mathcal{D} |\mathcal{M}_i)\equiv Z_i$ is the conditional probability that the data $\mathcal{D}$ is observed when a model $\mathcal{M}_i$ is given, referred to as evidence of a model $\mathcal M_i$.
Assuming that all the prior probabilities are the same, i.e., $P(\mathcal{M}_i) = P(\mathcal{M}_j)$ for all $(i,~j)$ pairs, then the probability that the model $\mathcal{M}_i$ is true among $N$ possible models is given by the ratio of the model evidence:
\begin{equation}
    \frac{P(\mathcal{M}_i|\mathcal{D})}{\sum_{j=1}^{N} P(\mathcal{M}_j|\mathcal{D})}
    = \frac{P(\mathcal{D}|\mathcal{M}_i)}{\sum_{j=1}^N P(\mathcal{D}|\mathcal{M}_j)} = \frac{Z_i}{\sum_{j=1}^N Z_j}.
\end{equation}
This is the key idea in the Bayesian approach towards model inference. The best-fit model can be found via the comparison of the model evidences.

To calculate the model evidence, we define the likelihood function $\mathcal{L}_{i}(\vec{\theta}_i) \equiv P(\mathcal{D} | \vec{\theta}_i, \mathcal{M}_i)$ where $\vec{\theta}_i$ is the parameter set necessary to specify the model $\mathcal{M}_i$. The model evidence (marginal likelihood) is then evaluated by marginalizing the likelihood function $\mathcal{L}_{i}(\vec{\theta}_i)$ over the parameter space $\Theta$ with a weight $\pi(\vec{\theta}_i)$:
\begin{equation} \label{eq:evidence}
    Z_i = P(\mathcal{D}|\mathcal{M}_i) = \int_\Theta P(\mathcal{D}|\vec{\theta}_i, \mathcal{M}_i) \pi(\vec{\theta}_i) d\vec{\theta}_i.
\end{equation}
Here, $\pi(\vec{\theta}_i) \equiv p(\vec{\theta}_i | \mathcal{M}_i)$ is a prior probability for the parameter set $\vec{\theta}_i$ given based on additional information or intuition~\cite{gelman2013}.

The Bayesian inference approach is effectively used to find the best parameter set for a given data set if a specific model is given or the best model is determined from the above model inference. 
The posterior probability distribution of the parameters is given by
\begin{equation} \label{eq:posterior}
    P(\vec{\theta}_i | \mathcal{M}_i, \mathcal{D}) = \frac{P(\mathcal{D} | \vec{\theta}_i, \mathcal{M}_i) P(\vec{\theta}_i|\mathcal{M}_i)}{P(\mathcal{D}|\mathcal{M}_i)} = \frac{\mathcal{L}_i(\vec{\theta}_i)\pi(\vec{\theta}_i)}{Z_i},
\end{equation}
and we report the medians of this distribution as the estimated values of the parameters
of a given model.
It is computationally nontrivial to perform the marginalization of the likelihood function [Eq.~(\ref{eq:evidence})] and to sample over the posterior distribution (\ref{eq:posterior}). For this task, here we use the method of annealed importance sampling~\cite{neal2001}. For further information on this method, see our MATLAB code for annealed importance sampling in the AnDi challenge GitHub~\cite{BIT}.

\subsection{The discrete-time L\'evy walk}
\label{ss:levywalk}
In the L\'evy walk model, a flight time $\tau$ of a ballistic run with speed $v$ and its step length $\Delta x$ is given by the joint probability 
\begin{equation}
\Psi(\Delta x, \tau) 
= \frac{1}{2}\psi(\tau)\delta (|\Delta x| - v\tau).
\end{equation}
Because of the constraint between flight time and step length, the whole dynamics are solely determined by characteristics of the flight time distribution $\psi(\tau)$. In the L\'evy walk we consider the class of $\psi(\tau)$ that asymptotically behaves as a power-law function at large times
\begin{equation}\label{eq:powerlaw}
\psi(\tau) \propto
\frac{1}{\tau^{1+\gamma}}
\end{equation}
with a power-law exponent $\gamma$ (i.e., the stability index of a L\'evy distribution) of $0<\gamma\leq 2$. The L\'evy walk has distinct diffusion dynamics depending on $\gamma$~\cite{zaburdaev2015}: the diffusion becomes ballistic for $0<\gamma<1$, sub-ballistic superdiffusive for $1<\gamma<2$, and diffusive for $\gamma\geq 2$. 
In this work, we are interested in the sub-ballistic L\'evy walk where the MSD increases with time as~\cite{zaburdaev2015}
\begin{equation}
\langle x^2(t) \rangle \propto t^{3-\gamma}~\hbox{at}~t\to\infty.    
\end{equation}
Thus, in this range of $\gamma$, the index of flight-time distribution [Eq.~(\ref{eq:powerlaw})] is related to the anomaly exponent via
\begin{equation}\label{eq:alphagamma}
\alpha=3-\gamma~\hbox{with}~1<\alpha<2.  
\end{equation}
For simplicity, hereafter, we denote the flight-time distribution $\psi(t)$ in terms of $\alpha$ [via Eq.~(\ref{eq:alphagamma})] throughout the text. 



While the L\'evy walk is defined with $\psi(\tau)$ in the continuous time-domain of $\tau>0$, time series in the experiment, such as particle trajectories obtained from a single-particle tracking experiment, are usually given in unit of a time lag that is bound by the time resolution of a measurement. To apply our Bayesian inference approach to such data, we here introduce a discrete-time L\'evy walk. Practically, the L\'evy walk trajectory is often generated in this way. The anomalous diffusion simulation package (the AnDi package~\cite{andipackage, andi_expl}) used in this work provides the discrete-time L\'evy walk.

\begin{figure}
\centering
\includegraphics[width=1\textwidth]{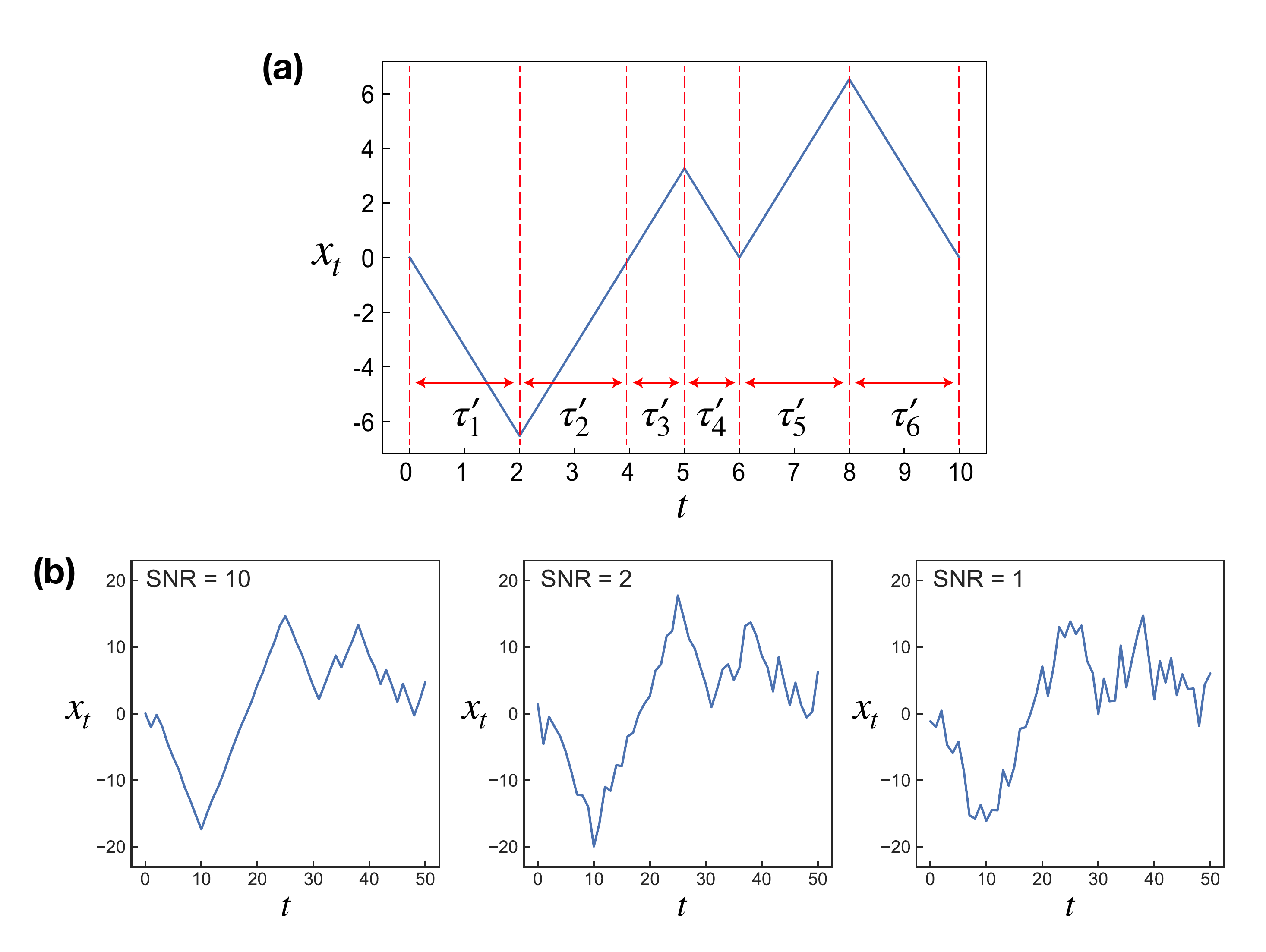} 
\caption{Sample trajectories for one-dimensional discrete-time L\'evy walks. (a) A L\'evy walk comprises ballistic flights whose flight events are specified by (velocity, flight time)=$\{(v_i=\pm v, \tau_i)\}$. The flight times $\tau\in(0,\infty)$ are randomly given by a power-law PDF $\psi(\tau)\propto \tau^{\alpha-4}$ ($1\leq \alpha \leq 2$). In the discrete-time L\'evy walk (shown in the panel), flight times are integers $\tau'(\in \mathbb{N})$ whose probability mass function is given by Eq.~(\ref{eq:andipl2}).  (b) Noisy discrete-time L\'evy walk trajectories simulated from the governing equation (\ref{eq:position}) with $\alpha = 1.36$. From Left to Right, the trajectories have a noise level with $\mathrm{SNR}=10, 2,$ and 1. See the text for further information. The trajectories are generated by the AnDi package~\cite{andipackage}. }
\label{fig:trj}
\end{figure}

In the discrete-time L\'evy walk, the flight times are integers, sampled by the following two steps. (1) A continuous random time $\tau(>0)$ is sampled by the transformation \begin{equation}\label{eq:andipl}
\tau \leftarrow F^{-1} (u;\alpha) = (1-u)^{-\frac{1}{3-\alpha}}    
\end{equation}
using a uniform random number $u$ in $(0,1)$
and $F(t) = \int_1^t \psi(\tau) d\tau =(3-\alpha) \int_1^t  \tau^{\alpha-4} d\tau$.
The normalized probability density function for $\tau$ is
\begin{equation} \label{eq:exactcont}
    \psi_\mathrm{c}^{\mathrm{target}}(\tau; \alpha) = \frac{3-\alpha}{\tau^{4-\alpha}} \quad \hbox{for~}\tau \geq 1.
\end{equation}
(2) The integer part of $\tau$ is taken using floor function $\tau' = \lfloor \tau \rfloor$. After the rounding-off, the probability mass function for $\tau'$ is given by
\begin{equation} \label{eq:andipl2}
\begin{aligned}
    \psi_\mathrm{d}^{\mathrm{target}}(\tau'; \alpha) &= \int_{\tau'}^{\tau'+1} \psi_\mathrm{c}^\mathrm{target}(\tau;\alpha) d\tau \quad \hbox{for}\,\,\tau'\in \mathbb{N}\\
    &= \frac{1}{\tau'^{3-\alpha}} - \frac{1}{(\tau'+1)^{3-\alpha}} \propto \frac{1}{\tau'^{4-\alpha}}.
\end{aligned}
\end{equation}
From now on, we call Eq.~(\ref{eq:exactcont}) the continuous target distribution and Eq.~(\ref{eq:andipl2}) the discrete target distribution. With the integer random times given by $\psi^\mathrm{target}_\mathrm{d}(\tau')$, we can generate a discrete-time L\'evy walk trajectory parametrized by $\vec{\theta} = \left \{ v, \alpha, \sigma_\mathrm{noise} \right \}$ via the governing equation
\begin{equation} \label{eq:position}
\begin{aligned}
    x_t &= x_t^\mathrm{LW} + \xi_t \\
    &= \left [ \sum_{i=1}^{N} (-1)^{r_i}v\tau'_i + (-1)^{r_{N+1}} v\tau'_\mathrm{res} \right ] +\xi_t
\end{aligned}
\end{equation}
with
\begin{equation}
    t= \sum_{i=1}^N \tau'_i + \tau'_\mathrm{res}.
\end{equation}
Figure~\ref{fig:trj}(a) shows a schematic description of our discrete-time L\'evy walk. For given observation time $t$, the walker (or particle) has $N$ complete flight events and an incomplete flight event. The duration time for the latter is described by a residual time $\tau'_\mathrm{res}= t-\sum_{i=1}^{N}\tau'_{i}$. The velocity of the walker during the $i$-th flight is $(-1)^{r_i}v$ where $r_i=0$ or $1$ with an equal probability. Additionally, we consider a L\'evy walk in a noisy environment. Physically, the noise may originate from localization errors during the tracking measurement or the thermal random agitation in the heat bath. Thus, in a complete picture, we invent the so-called \emph{noisy} L\'evy walk trajectory by superimposing a Gaussian random noise $\xi_t$ to the noise-free trajectory in Eq.~(\ref{eq:position}). The Gaussian noise is characterized by $\xi_t\sim \mathcal N(0,\sigma_\mathrm{noise}^2)$ where $v/\sigma_\mathrm{noise}$ (or $\sigma_D/\sigma_\mathrm{noise}$ after standardizing the trajectory in the later section) is understood as the signal-to-noise (SNR) ratio. Figure~\ref{fig:trj}(b) shows examples of the noisy L\'evy walk trajectories for varying SNRs. When the noise level is significantly high ($\mathrm{SNR}=1$), the noisy L\'evy walk no longer looks like a L\'evy walk.   


\section{Hidden Markov model for L\'evy walks \& Bayesian inference} \label{sec:modeling}

\subsection{Hidden Markov model for L\'evy walks } \label{sec:sec32}

In this section, we develop our theoretical formalism of the Bayesian inference framework for L\'evy walks. For this, we first build up a solvable model that appropriately describes the discrete-time L\'evy walk process in the scheme of the Bayesian formalism. Our approach is to establish a hidden Markov model for non-Markovian processes governed by a power-law relaxation dynamics using the Markovian decomposition method~\cite{goychuk2009}. 

Now let us consider a flight-time distribution with a power-law decay up to a time $T$, which is referred to as a cutoff time and usually considered the maximum observation time window in the measurement. Such power-law distributions can be represented by the superposition of a finite number of independent exponential distributions (i.e., Ornstein-Uhlenbeck processes); see the method developed in Ref.~\cite{goychuk2009}. In this approximation scheme (with a slight modification), the continuous target distribution can be expanded in terms of multiple exponential distributions as such: 
\begin{equation} \label{eq:approxtarget}
    \psi_\mathrm{c}^{\mathrm{target}}(\tau;\alpha)\approx \sum_{i=1}^{N_k} P_{i; \alpha} k_i e^{-k_i(\tau-1)} \qquad (\tau \geq 1)
\end{equation}
where the inverse of the relaxation time of the $i$-th component and its statistical weight are, respectively, given by
\begin{align}
    k_i &= \frac{k_{\mathrm{fast}}}{b^i}\qquad (i=1, 2, \cdots, N_k) \label{eq:rate}\\
    P_{i;\alpha} &= \frac{k_i^{3-\alpha}e^{-k_i}}{\sum_{j=1}^{N_k} k_j^{3-\alpha}e^{-k_j}} \label{eq:transition}.
\end{align}
See the \ref{sec:powerlaw_deriv} for the validation of Eq.~(\ref{eq:approxtarget}). Here, $b$ is a scale parameter, $k_{\mathrm{fast}}$ is the high-frequency cutoff, and $N_k$ determines the long-time exponential cutoff $T\approx (4-\alpha) b^{N_k}/k_{\mathrm{fast}}$. We set $b = 4$, $k_\mathrm{fast} = 8$, and $N_k = 5$ through the work. In this scheme, we regard the power-law distributed flight times (within the observed time window) as the inter-arrival times among $N_k$-component independent Markov processes. The sojourn time in the state $s_i$ has the probability density function
\begin{equation}\label{eq:interarrival}
\phi_i(\tau) = k_i e^{-k_i (\tau-1)}.
\end{equation}
The state transition from state $s_i$ to state $s_j$ occurs with a probability $P_{j;\alpha}$ when the process leaves the state $s_i$. This process gives the inter-arrival time distribution Eq.~(\ref{eq:approxtarget}) in the form of $\sum_{i=1}^{N_k} P_{i;\alpha}\phi_i(\tau)$.

\begin{figure}
\centering
\includegraphics[width=1\textwidth]{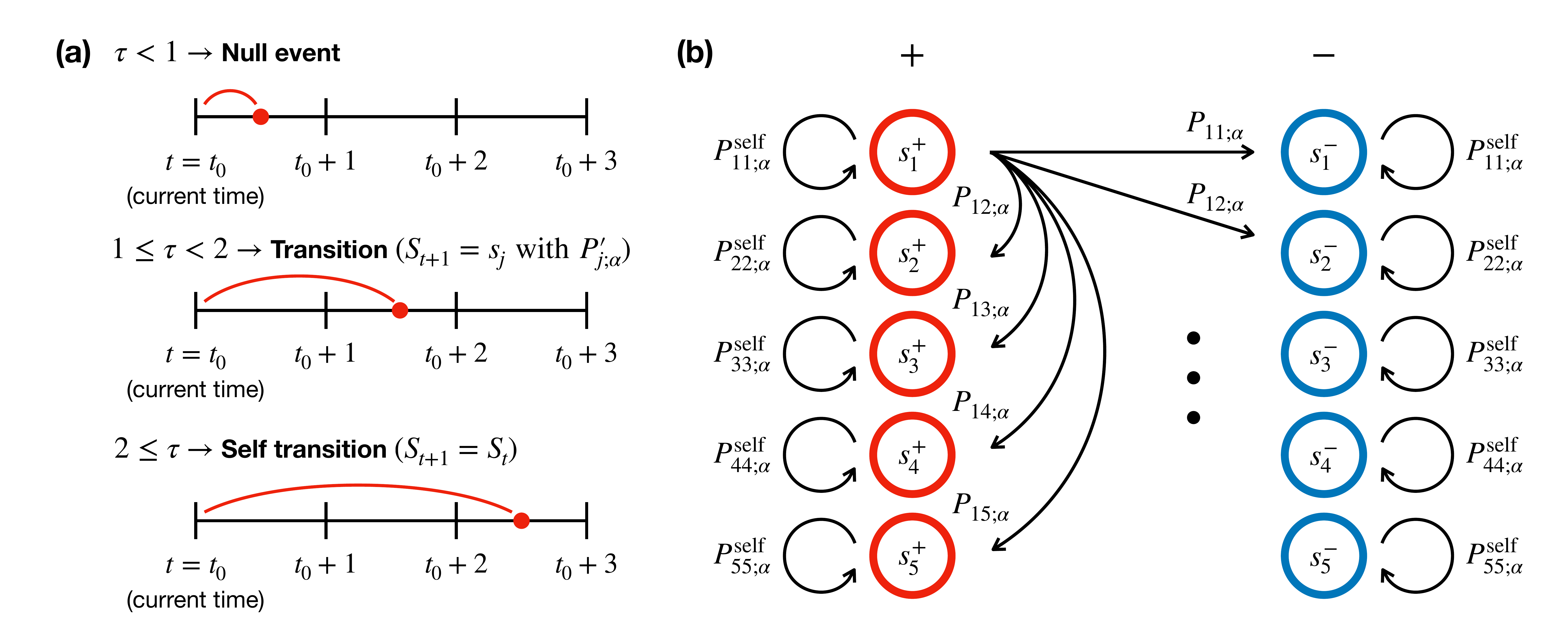}
\caption{(a) Transition strategy of a random walker in our Markovian approximation of discrete-time L\'evy walk model. A walker in a state $s_i^\pm$ samples transition time $\tau$ from $\phi_i(\tau)$ every time step. The probability of $\tau<1$ event~(top) is zero. If a walker samples $1 \leq \tau < 2$~(middle), it changes the state from $s_i^\pm$ to $s_j^\pm$ with probability $P_{j;\alpha}$ ($\frac{1}{2}P_{j;\alpha}$ for each of the two states $s_j^+$ and $s_j^-$). If a walker chooses $2\leq \tau$~(bottom), it maintains its original state ($S_{t+1} = S_{t}$). (b) Schematic illustration of hidden Markov model for one-dimensional L\'evy walk. In this model, the `$+$' states~(red circles) have positive velocity $+v$ and `$-$' states~(blue circles) have negative velocity $-v$. The probabilities of self-transitions and transitions for every time step are denoted by $P_{ii;\alpha}^\mathrm{self}$ and $P_{ij;\alpha}$, which are given by (\ref{eq:selfprob}) and (\ref{eq:transprob}), respectively.}
\label{fig:hmmscheme}
\end{figure}

Now we set up the proper rounding-off process to make discrete-time Markov chains of $N_k$ components having the inter-arrival times governed by $\psi^\mathrm{target}_d(\tau')$. The walker's state is renewed at every time unit such that it is changed to one of the other states (\emph{transition}) or maintains the same state (\emph{self-transition}).
Figure~\ref{fig:hmmscheme}(a) illustrates how the transition and self-transition events are determined from the continuous inter-arrival times. Let us assume that at $t=t_0$ the walker (Markov chain) is in the state $S_{t_0}=s_i$ and samples the transition time $\tau$ according to $\phi_i(\tau)$. Here $S_t$ denotes the walker's state at time $t$. (i) If $1\leq \tau<2$, it is allocated to $\tau'=1$ from the rounding-off process. Thus, the walker in the next time ($t=t_0+1$) should change its state to a new one, say, $S_{t_0+1}=s_j$. In this case, the state transition occurs with the transition probability  
\begin{equation}
    \left [ \int_1^2 \phi_i(\tau) d\tau \right ] P_{j; \alpha} = \left ( 1-e^{-k_i}\right ) P_{j; \alpha.}
\end{equation}
(ii) If $\tau\geq 2$, then $\tau'\geq 2$ and the walker stays in the same state at $t=t_0+1$ ($S_{t_0+1} = S_{t_0} = s_i$). This is the case called the self-transition, which occurs with the transition probability 
\begin{equation}
    \int_2^\infty \phi_i(\tau) d\tau = e^{-k_i}.
\end{equation}
(Beware that the event of $\tau<1$ is the null case because such events are not allowed from $\phi(\tau)$).  
The probability that a walker in the $i$-th state has an integer sojourn time $\tau'$ is
\begin{equation}
    \int_{\tau'}^{\tau'+1} \phi_i(\tau) d\tau = e^{-k_i(\tau'-1)}-e^{-k_i\tau'}, \label{eq:atonce}
\end{equation}
which can be understood in the Markov chain description as the probability that the walker maintains the same state $(\tau'-1)$ times and then changes its state, i.e., 
\begin{equation}
    (e^{-k_i})^{\tau'-1}\cdot(1-e^{-k_i}) = e^{-k_i(\tau'-1)} - e^{-k_i \tau'}. \label{eq:multistep}
\end{equation}
Therefore, by simply repeating the above process every time step, we obtain the process having integer inter-arrival times of $\tau'\geq 1$ with the distribution
\begin{equation}
\begin{aligned}
    \sum_{i=1}^{N_k} P_{i; \alpha} \int_{\tau'}^{\tau'+1} \phi_i(\tau) d\tau &= \int_{\tau'}^{\tau'+1} \sum_{i=1}^{N_k} P_{i; \alpha} \phi_i(\tau) d\tau  \quad (\tau' \in \mathbb{N}) \\
    &\approx \int_{\tau'}^{\tau'+1} \psi^\mathrm{target}_\mathrm{c} (\tau;\alpha) d\tau \\
    &= \psi^\mathrm{target}_\mathrm{d}(\tau';\alpha).
\end{aligned}
\end{equation}

Figure~\ref{fig:hmmscheme}B illustrates our Markov chain scheme based on the above formalism. Here we incorporate the fact that the L\'evy walk (in 1D) has two velocity states, $\pm v$, for given flight time. This is described in our model such that the state $s_i$ has two substates $s_i^+$ (for $+v$) and $s_i^-$ (for $-v$) with the same transition rate $k_i$. Thus, there are $2N_k$ states in total in the model and the transition probability from the $i$-th state to the $j$-th state is given by
\begin{equation}\label{eq:transprob}
    P_{ij;\alpha} = \frac{(1-e^{-k_i})\cdot P_{j; \alpha}}{2}.
\end{equation}
As shown in Fig.~\ref{fig:hmmscheme}(b), the change of the velocity state from $s_i^+$($s_i^-$) to $s_i^-$($s_i^+$) is considered as a transition with the probability $P_{ii;\alpha}$. On the other hand, if the process changes its state from $s_i^+(s_i^-)$ to $s_i^+(s_i^-)$, this is effectively a self-transition. The effective probability of self-transition is then given by
\begin{equation}\label{eq:selfprob}
    P_{ii;\alpha}^{\mathrm{self}} = e^{-k_i} + \frac{(1-e^{-k_i})\cdot P_{i; \alpha}}{2}.
\end{equation}
We numerically confirm our hidden Markov model for discrete-time L\'evy walks in Sec.~\ref{ss:HMMresult}.

\subsection{The likelihood function} \label{ss:likelihood}

The above discrete-time Markov chain describes the dynamics of the latent states $S_t$. For given parameter set $\vec{\theta} = \{ \alpha, v, \sigma_\mathrm{noise} \}$ and $S_t$, the conditional probability of observing $\Delta x_t = x_{t+1} - x_t$, i.e., the so-called emission probability, is given by
\begin{equation} \label{eq:emission}
    P(\Delta x_t | \mathcal M_\mathrm{L\acute{e}vy}, S_t, \vec{\theta}) = \frac{1}{\sqrt{4\pi \sigma_\mathrm{noise}^2}} \exp\left (-\frac{1}{2}\left ( \frac{\Delta x_t \pm v}{\sqrt{2}\sigma_\mathrm{noise}}\right )^2 \right ).
\end{equation}
In this expression, $\Delta x_t - v$ corresponds to the state $S_t = s_i^+$ ($i=1,\ldots,N_k$) and $\Delta x_t + v$ to the state $S_t = s_i^-$, respectively. The factor $\sqrt{2}\sigma_\mathrm{noise}$ is attributed to the fact that the displacement is $\Delta x_t = x_{t+1}^\mathrm{LW} -x_t^\mathrm{LW} + \xi'_{t}$ where $\xi'_t$ is a Gaussian random noise of $\mathcal{N}(0, 2\sigma^2_\mathrm{noise})$. We consider this term as an approximate way to include measurement noise for the positions, i.e., we should really have had $\xi'_t = \xi_{t+1} - \xi_{t}$ with the $\xi_{t}$ independent. This approximation ignores the correlations between steps that measurement noise induces, and we will see later that it does not work well when $\sigma_\mathrm{noise}$ is comparable to the Levy walk step size, i.e., when $\mathrm{SNR}$ is not larger than unity. 
For brevity, below we omit the notation of $\mathcal M_\mathrm{L\acute{e}vy}$ in the conditional probabilities.

In the discrete-time L\'evy walk, a trajectory $x_t$ can be represented by a series of the unit-time displacements $\mathcal D = \{ \Delta x_1, \Delta x_2, \cdots, \Delta x_T\}$. Thus, when the latent state is given by $\mathcal S = \left \{ S_1, S_2, \cdots, S_T\right \}$, the probability of finding $\mathcal D$ (or $x_t$) is written as the product of the emission probabilities
\begin{equation} \label{eq:seqL}
    P(\mathcal D |\mathcal S, \vec{\theta}) = P(\Delta x_1 | S_1, \vec{\theta})P(\Delta x_2 | S_2, \vec{\theta})\cdots P(\Delta x_T | S_T, \vec{\theta}).
\end{equation}
We should marginalize Eq.~(\ref{eq:seqL}) over all possible sequences of the latent state $\mathcal S = \{S_1, S_2, \cdots, S_T\}$ to obtain the likelihood function
\begin{equation}\label{eq:likehoodfunc}
    P(\mathcal D | \vec{\theta}) = \sum_\mathcal{S} P(\mathcal D | \mathcal S, \vec{\theta}) P(\mathcal S | \vec{\theta}).
\end{equation}
In this work, we use the forward algorithm method to calculate Eq.~(\ref{eq:likehoodfunc}) for obtaining the corresponding likelihood function~\cite{Rabiner1989}. The summation over $\mathcal S$ is carried out in an iterative way with the help of the probability function
\begin{equation} \label{eq:alphadef}
    \beta(S_t = s_i^\pm) \equiv P(\Delta x_1, \cdots, \Delta x_t, S_t = s_i^\pm | \vec{\theta}).
\end{equation}
This probability function satisfies the recurrence relation
\begin{equation}\label{eq:recurrence}
\begin{aligned}
        \beta(S_{t+1} = s_i^\pm) &= P(\Delta x_1, \cdots, \Delta x_t, \Delta x_{t+1}, S_{t+1} = s_i^\pm | \vec{\theta})\\
        &= P(\Delta x_{t+1} | S_{t+1} = s_i^\pm, \vec{\theta}) \\
        &\quad \times \sum_{j=1}^{N_k} \left [ \beta(S_{t} = s_j^+) P(S_{t+1} = s_i^\pm | S_t = s_j^+) \right. \\
        &\qquad \qquad+ \left. \beta(S_{t} = s_j^-) P(S_{t+1} = s_i^\pm | S_t = s_j^-)\right ].
\end{aligned}
\end{equation}
In this expression, we figure that $P(\Delta x_t | S_t = s_i^\pm, \vec{\theta})$ is the emission probability~(\ref{eq:emission}) and $P(S_{t+1} = s_i^\pm | S_t = s_j^\pm)$ is the transition probabilities between two distinct states [Eq.~(\ref{eq:transprob})] or between the same state (self-transition) [Eq.~(\ref{eq:selfprob})].
The initial condition of the recurrence relation is
\begin{equation} \label{eq:forward}
\begin{aligned}
    \beta(S_1 = s_i^\pm) &= P(\Delta x_1, S_1 = s_i^\pm | \vec{\theta}) \\
    &= P(\Delta x_1 | S_1 = s_i^\pm, \vec{\theta}) P(S_1 = s_i^\pm).
\end{aligned}
\end{equation}
The initial distribution of the states $P(S_1=s_i^\pm)$ is given by Eq.~(\ref{eq:transition}) provided that the process starts at $t=0$ (or more generally, the initial time $t=0$ is the renewal time). Alternatively, if the process is observed after a long time where it has become stationary, then an extra factor of the average length of the time intervals, $C/(1-e^{-k_i})$, should be multiplied on the right hand side of Eq.~(\ref{eq:transition}) to obtain the initial distribution. $C$ is a normalizing constant of the distribution.

The recurrence relation Eq.~(\ref{eq:forward}) is given in terms of the transition probability, emission probability, and the initial distribution of the states that can be computable. Therefore, starting from Eq.~(\ref{eq:forward}), we can iteratively obtain $\beta(S_T=s_i^\pm) = P(\Delta x_1, \cdots, \Delta x_T, S_T = s_i^\pm | \vec{\theta})$. The likelihood function is then simply obtained by marginalizing $\beta(S_T = s_i^\pm)$ over all possible states:
\begin{equation}\label{eq:likelihood}
    P(\mathcal D | \vec{\theta}) = P(\Delta x_1, \cdots, \Delta x_T | \vec{\theta}) = \sum_{i=1}^{N_k} \left [ \beta(S_T = s_i^+) + \beta(S_T = s_i^-) \right ].
\end{equation}

Note that changes in direction and observations of the L\'evy walker can only happen at integer values of time. In case observations happen with a camera that is operated independently of the system it observes, this can be considered an approximation that is valid in the limit of the camera frame rate being much faster than the typical interval between changes of direction.

\subsection{Prior distributions}
For Bayesian inference, we need to specify the prior distributions for parameters $\vec{\theta} = \{ \alpha, v, \sigma_\mathrm{noise}\}$. We use a uniform prior distribution for the anomaly exponent $\alpha$ based on the fact that our L\'evy walk describes a superdiffusive and Fickian dynamics with the anomaly exponent in the range of $1\leq \alpha\leq 2$. The prior distribution is set to be
$P(\alpha) = 1$ for $1\leq \alpha \leq 2$ or zero otherwise. 
For the walker's speed $v(>0)$, we use a folded Gaussian distribution $P(v) = \sqrt{\frac{2}{\pi}}\exp\left (-\frac{v^2}{2}\right )$ in the positive regime. We use the Gaussian distribution because the trajectories we analyze in Sec.~\ref{sec:result} have a standard normal distribution of speed $v$ after standardization. The pre-processing (standardization) of the trajectories is further explained in Sec.~\ref{sec:paramest}. In the case of the background noise strength $\sigma_\mathrm{noise}$, its magnitude compared to the signal can be arbitrary. Here, we use a folded Cauchy distribution $P(\sigma_\mathrm{noise}) = \frac{2}{\pi \left ( 1 + \sigma_\mathrm{noise}^2 \right )}$ with $\sigma_\mathrm{noise} > 0$ as a prior distribution. The Cauchy distribution is a heavy-tailed distribution, allowing $\sigma_\mathrm{noise}$ to be an arbitrarily large value.


\section{Results I: Numerical implementation of the Bayesian inference and parameter estimation} \label{sec:result}
\subsection{Numerical implementation of the hidden Markov model} \label{ss:HMMresult}
\begin{figure}
\centering
\includegraphics[width=1\textwidth]{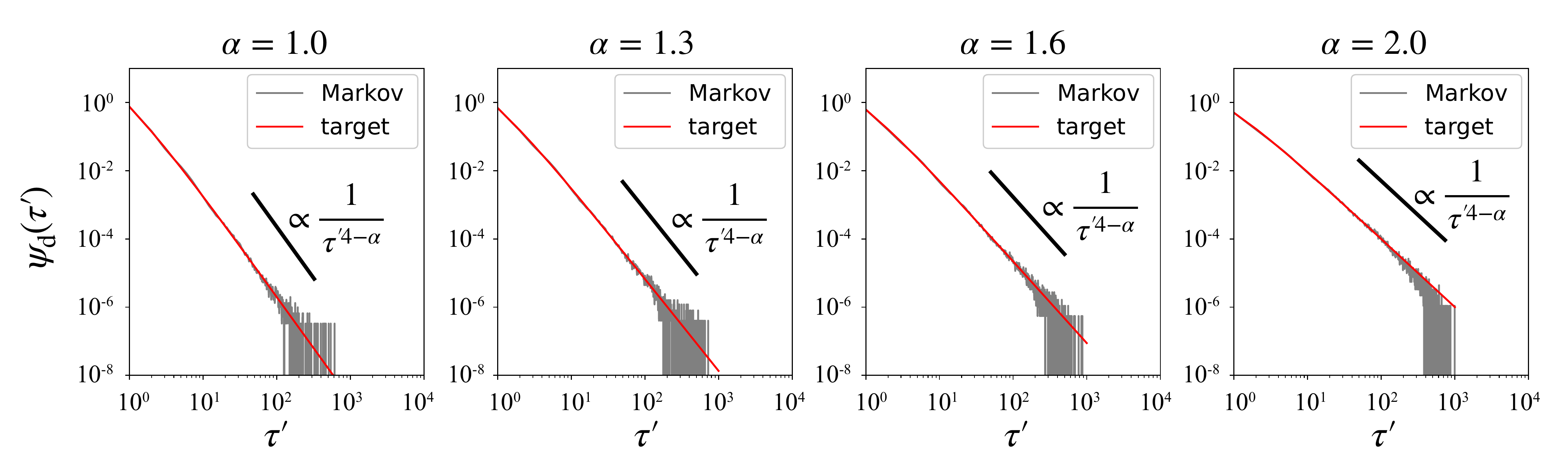}
\caption{Flight time statistics sampled from the inter-arrival times for our Markov chain model (gray lines). In the simulation of the Markov chain model we use the parameters $N_k=5$, $b = 4$,  $k_{\mathrm{fast}} = 8$, and $\alpha = 1.0,\,1.3,\,1.6,$ and $2.0$ (from Left to Right). The red lines show the discrete target distribution $\psi^\mathrm{target}_d (\tau')$ shown in Eq.~(\ref{eq:andipl2}). }
\label{fig:distribution}
\end{figure}

We have numerically implemented our hidden Markov model for the discrete-time L\'evy walk. In Fig.~\ref{fig:distribution} we plot the flight-time distributions obtained from the simulated inter-arrival times of the Markov chain at $\alpha = 1.0,\,1.3,\,1.6$, and $2.0$.
We allocate 5000 random walkers on every Markov chain~(Fig.~\ref{fig:hmmscheme}(b)) and collect the inter-arrival times for the transition events between distinct latent states up to $t=1000$. For all cases, we find the numerically obtained inter-arrival times~(gray lines, Fig.~\ref{fig:distribution}) are in excellent agreement with the expected power-law target distribution~(\ref{eq:andipl2})~(red lines). While the longest exponential cutoff is expected to start at $T\approx O(300)$, the power-law scaling in the numerical data seems to extend longer than $T$. The tail of the distribution for $t \gtrsim T$ is noisy because this part is sampled from the rare events of the $N_k$-th latent state. The exponential cut-off time $T$ gets shorter as $\alpha$ increases. For $\alpha = 2$, the distribution shows the expected power-law up to $T\approx 256$, and then it is slightly off the power-law due to the exponential cutoff. 

\begin{figure}
\centering
\includegraphics[width=1\textwidth]{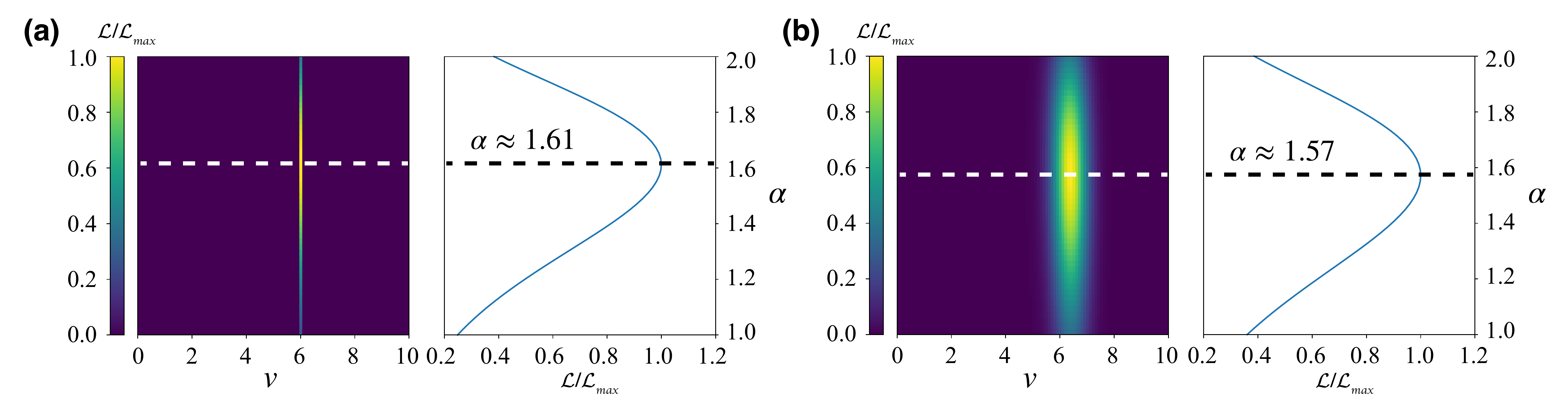}
\caption{Heatmap of likelihood functions for a L\'evy walk trajectory with $\alpha = 1.6$ and the speed $v\approx 5.92$. (a) The likelihood function for the L\'evy trajectory without the noise ($\sigma_\mathrm{noise}= 0$). (b) The likelihood function for the trajectory with $\sigma_\mathrm{noise} = 3$. In (a) \& (b), the heatmap shows the likelihood function marginalized over $\sigma_\mathrm{noise}$ as a function of $v$ and $\alpha$. The line plot represents the likelihood function marginalized over $\sigma_\mathrm{noise}$ and $v$, where the dashed line shows the $\alpha$ at which the likelihood function gives the maximum value.
}
\label{fig:likelihood}
\end{figure}

We numerically obtain the likelihood function by the iterative method [Eq.~(\ref{eq:likelihood})] and confirm the validity of our method. For this purpose, we generate L\'evy walk trajectories of length $100$ time steps with the flight-time exponent $\alpha = 1.6$ and $v \approx 5.92$ (i) without the background noise and (ii) with the noise with $\mathrm{SNR}=2$. We estimate the likelihood function for the trajectories over the parameter space $v \in [0, 10]$, $\alpha \in [1, 2]$, and $\sigma_\mathrm{noise} \in [0.01, 5]$. Figure~\ref{fig:likelihood} visualizes the numerically estimated likelihood functions. In the figure, the heatmaps show the likelihood functions as a function of $v$ and $\alpha$ that are marginalized over the background noise $\sigma_\mathrm{noise}$; the line plots are the likelihood function as a function of $\alpha$ marginalized over both $\sigma_\mathrm{noise}$ and $v$. All the likelihood functions are divided by their maximum value $\mathcal{L}_\mathrm{max}$ for normalization. (1) The noise-free L\'evy walk trajectory  [Fig.~\ref{fig:likelihood}(a)]. When such ideal L\'evy walk data are given, the likelihood function expectedly produces the maximum probability at the parameter ($v$, $\alpha$) that is very close to the true value. It is noted that the shape of the likelihood function around the maximum is dependent on the parameter. While the likelihood function appears to be a sharply peaked function for $v$, it is a broad distribution for $\alpha$ (see the line plot). This is because there are much more samples (displacement vectors) for inferring $v$ than for inferring $\alpha$ that relies on fewer samples of flight times.
(2) The noisy L\'evy walk trajectory [Fig.~\ref{fig:likelihood}(b)]. We find that the noise makes the likelihood function dispersed in the parameter space $v$ while the profile of the likelihood function for $\alpha$ is almost unaffected by the noise. The increased uncertainty for $v$ is understandable. The speed of a flight can be fluctuating by the noise in the unit of time step. Despite the dispersion, the likelihood function still has the maximum at around the input speed. This suggests that the Bayesian approach can successfully estimate the true parameter value ($\alpha$ \& $v$) even from noisy trajectories.

\subsection{Parameter estimation via the Bayesian inference} \label{sec:paramest}
Based on the likelihood function constructed above, we have implemented a Bayesian inference method to estimate model parameters from the trajectory data. The L\'evy walk is characterized by three model parameters: the power-law exponent of flight-time distribution $\gamma=3-\alpha$ (or, equivalently, the anomaly exponent $\alpha$), the walker's flight speed $v$, and the background noise strength $\sigma_\mathrm{noise}$. In this work, we focus on inferring the power-law exponent of the flight-time statistics $\alpha$ for given L\'evy walk trajectories. The other parameters, such as $v$, can be easily inferred by available statistical approaches, e.g., by averaging absolute values of unit displacements.

For our study, we generate a total of 1100 L\'evy walk trajectories with various trajectory lengths $N_\mathrm{step} \in \{ 10, 100, 200, \cdots, 900, 1000\}$. The trajectories are produced by the AnDi package~\cite{andipackage}. The anomaly exponent $\alpha$ and the speed $v$ are randomly given such that  $\alpha \sim \mathrm{unif}(1, 2)$ and $v \sim \mathrm{unif}(0, 10)$, respectively. We standardize the trajectories using the protocol described in Ref.~\cite{andi2021}, in which the trajectories are regularized so that the standard deviation of the displacement distribution has $\sigma_D \sim \mathcal N (0, 1)$. This pre-processing is required to avoid the unexpected effect of step sizes on both the parameter estimation and model classification. We then add the three different levels of Gaussian noises to the 1100 trajectory with an $\mathrm{SNR} = \sigma_D / \sigma_\mathrm{noise} \in \{1, 2, 10\}$, thus in total generating 3300 noisy L\'evy trajectories.

\begin{figure}
\centering
\includegraphics[width=1\textwidth]{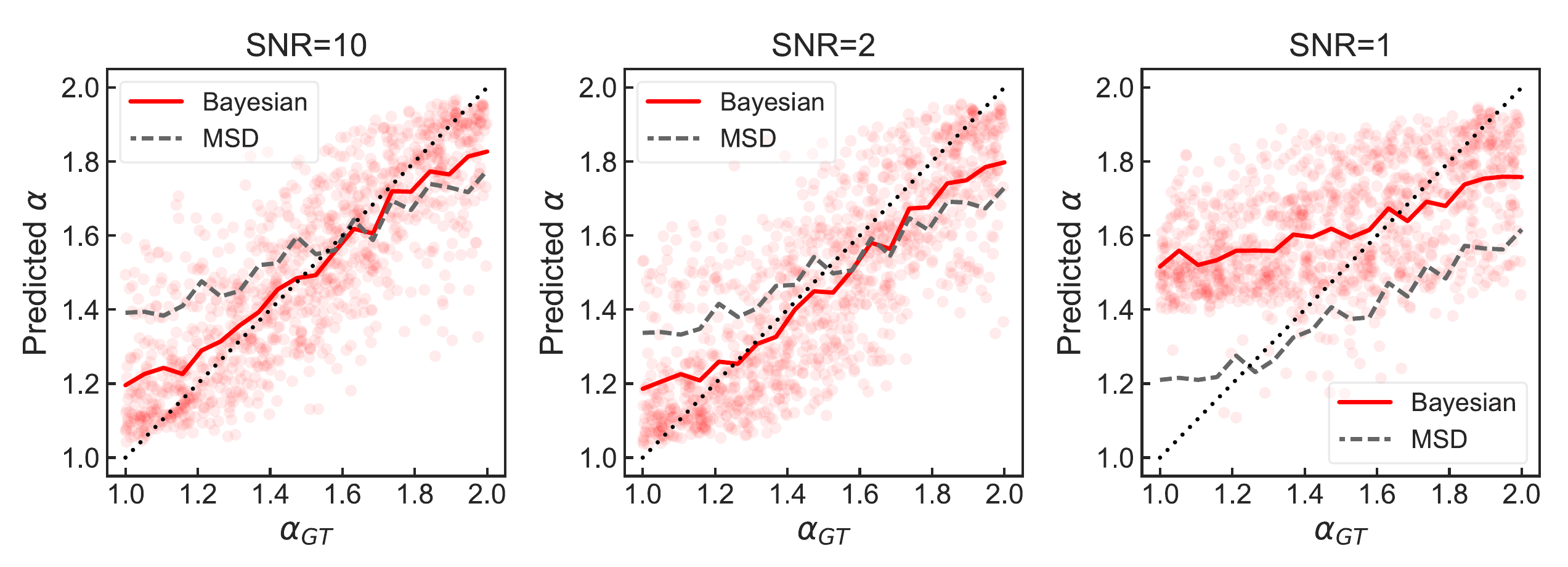}
\caption{The predicted $\alpha$ values from our Bayesian method vs the ground-true value $\alpha_\mathrm{GT}$. From Left to Right, the plot correspond to $\mathrm{SNR}=10,/2,$, and 1. 
The solid line (red) shows the average of the predicted $\alpha$ values for given $\alpha_\mathrm{GT}$. For reference, the dashed line depicts the corresponding curve for $\alpha$ estimated from the fitting of time-averaged MSD curves. We fit the linear part of the MSD curves in log-log plot, where the lag time $\Delta$ ranges from $2 \leq \Delta \leq 21$ for the trajectories whose lengths are over $100$. For the shorter ones, we took $2 \leq \Delta \leq 8$. The dotted line is the guide line of $y=x$. }
\label{fig:alphas}
\end{figure}

We perform our Bayesian inference analysis to these noisy L\'evy walk trajectories and estimate the anomaly exponent $\alpha$, more precisely, the power-law exponent of the flight-time distribution. In Fig.~\ref{fig:alphas}, we compare the predicted $\alpha$ to the ground-true value $\alpha_\mathrm{GT}$ used for generating the data. In each scatter plot, the solid line is the average of the predicted $\alpha$ for given $\alpha_\mathrm{GT}$. When the background noise is not significant ($\mathrm{SNR}=10,~2$), the predicted $\alpha$ (solid line) is in good agreement with $\alpha_\mathrm{GT}$. We confirm that the estimation of $\alpha$ via the Bayesian inference is more accurate than that from the fitting of the MSD curve (the dashed line). When the noise becomes comparable with the signal ($\mathrm{SNR}=1$), both methods show poor performance.

\begin{figure}
\centering
\includegraphics[width=1\textwidth]{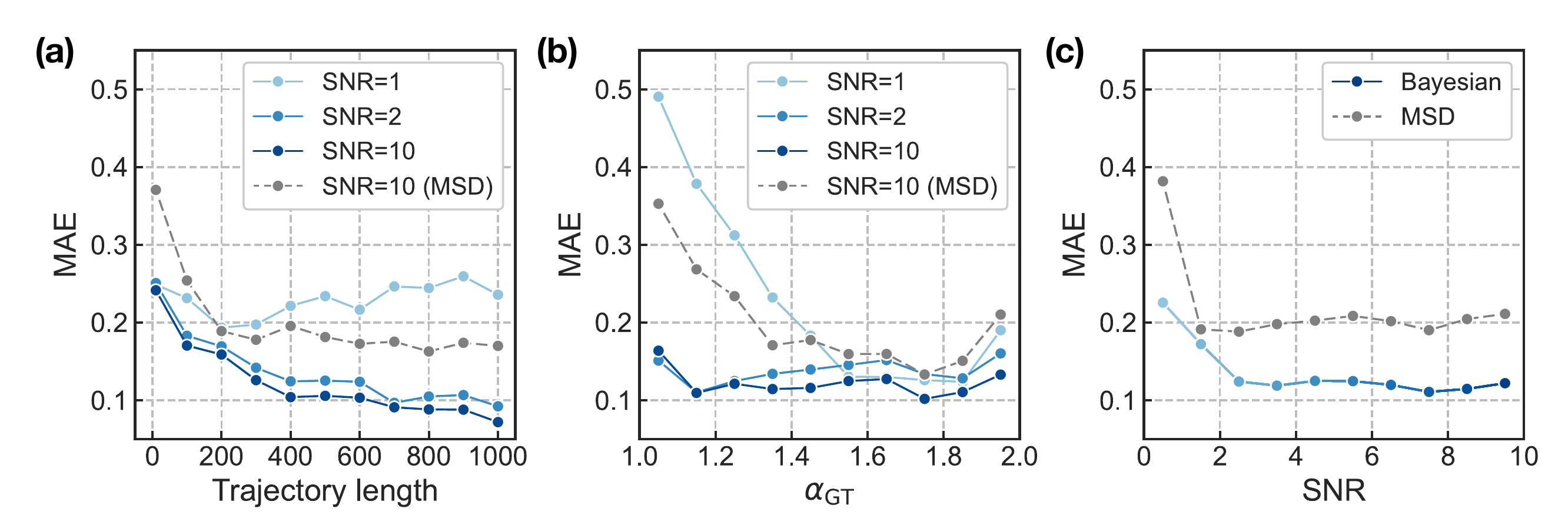}
\caption{Mean absolute error (MAE) from our Bayesian parameter estimation (line-dot) and from the fitting of MSD curves (dashed-dot). (a) MAE as a function of the trajectory length. (b) MAE as a function of $\alpha_{\mathrm{GT}}$. Each data point represents the average value for all trajectory lengths at a given $\alpha_{\mathrm{GT}}$ with half bin width $|\Delta \alpha_\mathrm{GT}| = 0.05$. (c) MAE as a function of SNR. Each data point represents the averaged AE for all trajectory lengths. For given $N_\mathrm{step}$, five hundred trajectories are generated with an SNR uniformly distributed over $(0, 10]$. In $x$-axis, half bin width is $|\Delta \mathrm{SNR}| = 0.5$.}
\label{fig:MAEs}
\end{figure}

For a more quantitative examination of the performance of our Bayesian inference method, we measure the absolute error~(AE), $|\alpha_i - \alpha_{\mathrm{GT}, i}|$, as a function of $N_\mathrm{step}$, $\alpha_\mathrm{GT}$, and SNR. In AE, $\alpha_i$ and $\alpha_{\mathrm{GT}}$ signify the predicted $\alpha$ and $\alpha_{\mathrm{GT}}$ for the $i$-th trajectory, respectively.
In Fig.~\ref{fig:MAEs}(a), we show the mean absolute error (MAE) as a function of $N_\mathrm{step}$. For $\mathrm{SNR}=10$ and $2$, the increase of trajectory length leads to more accurate estimation. For all trajectory lengths, the Bayesian method performs much better than the fitting from MSDs. For the case of $\mathrm{SNR}=1$, on the contrary, the inference goodness is not improved even if the length is increased. In this case, intense noises impede the accurate inference of flight times.

Figure~\ref{fig:MAEs}(b) shows the MAE from all trajectory lengths as a function of $\alpha_\mathrm{GT}$.
For $\mathrm{SNR}=10$ and $2$, the inference goodness is almost insensitive to $\alpha_\mathrm{GT}$ although the error seems to be relatively larger at $\alpha_\mathrm{GT}=1$ and $2$. Note that even at $\mathrm{SNR}=10$, the MAE from the fitting method (MSD) is considerably high for these cases.
When the noise level is too high ($\mathrm{SNR}=1$), the inference performance tends to be poorer as $\alpha_\mathrm{GT}$ is close to unity. This tendency occurs because the zigzag flight patterns frequently occurring in L\'evy walk trajectories with $\alpha_\mathrm{GT}\sim 1$ are easily confused with the noisy dynamics (when an SNR is small), so the short flight times cannot be correctly extracted. Consequently, as seen in Fig.~\ref{fig:alphas} ($\mathrm{SNR}=1$), the L\'evy walk trajectory with $\alpha_\mathrm{GT}\approx 1$ is inferred to the L\'evy walk with a predicted $\alpha>1$. 

In Fig.~\ref{fig:MAEs}(c) we examine the effect of SNR on MAE. The result shows that MAE abruptly decreases with increasing an SNR from $1$ to $\approx 2$, and after then it saturates for $\mathrm{SNR}\gtrsim 2$. This suggests that under a noisy environment the Bayesian inference method extracts the information of $\alpha$ (i.e., the power-law exponent in the flight-time distribution) with the accuracy at the noise-free condition. Surprisingly, such a robust performance is achievable against the noise level as significant as $\mathrm{SNR}\approx2$. We also notice that even in the strong noise condition ($\mathrm{SNR}\lesssim 1$), the Bayesian inference is better than the fitting of MSD method in the parameter estimation. 


\section{Results II: Model classification}\label{sec:results2}
In this section, we apply our Bayesian inference method for classifying models and comparing their likelihoods. Our task is to compare several diffusion models for a given trajectory in the scheme of Bayesian inference and find the most probable model in accordance with the data. For this task, we additionally consider two diffusion models---apart from the L\'evy walk (LW)---describing anomalous diffusion with $0< \alpha \leq2$. The two anomalous models are fractional Brownian motion~(FBM)~\cite{mandelbrot1968} and scaled Brownian motion~(SBM)~\cite{lim2002, jeon2014}. We refer to a recent review paper~\cite{metzler2014} for the mathematical introduction of these models and their applications to biology and various physical systems. Although the three models (LW, FBM, and SBM) can describe the superdiffusive process quantified by the MSD (\ref{eq:msd}), they are based on distinct theoretical context and physical mechanisms. Accordingly, their statistical characteristics are distinguished, enabling us to use the Bayesian approach for model inference.

For a numerical test, we generate trajectory data for the three diffusion models. The data preparation is the same as in Sec.~\ref{sec:paramest} for the noisy L\'evy walk trajectory. For each model, there are a total of 1100 trajectories with $N_\mathrm{step} \in \{10, 100, \cdots, 900, 1000\}$ (and 100 trajectories for given $N_\mathrm{step}$). 
The $\alpha$ is randomly sampled from $[1, 2]$ for the three diffusion models. All of these trajectories are standardized and superimposed with background noise in the same way described in Sec.~\ref{sec:paramest}.
The likelihood functions for FBM~\cite{krog2018} and SBM~\cite{samuSBM} are available from our previous work. We refer to Ref.~\cite{krog2018} for our Bayesian inference of FBM. Based on that, we have implemented the MATLAB code for the Bayesian model inference for LW, FBM, and SBM~\cite{BIT}.

\begin{figure}
\centering
\includegraphics[width=1\textwidth]{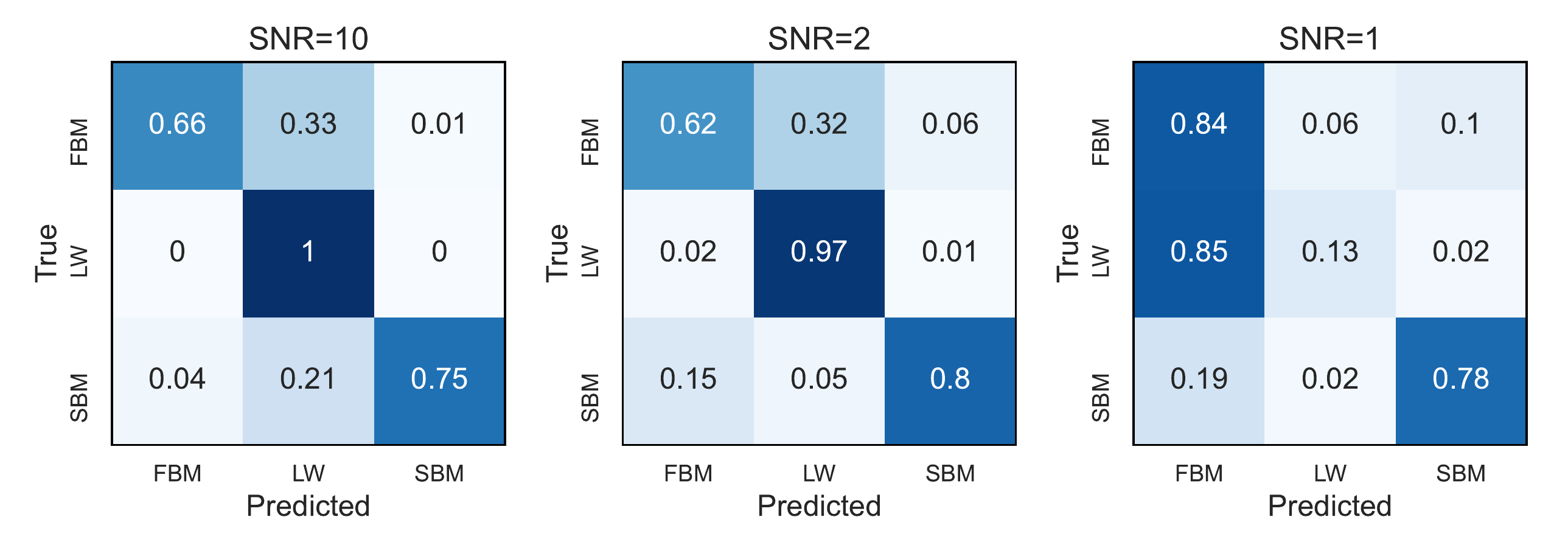}
\caption{Confusion matrices obtained from the Bayesian model classification. From Left to Right, the matrices show the results at $\mathrm{SNR}=10$, $2$, and $1$, respectively. The $(i,j)$-component in a matrix indicates the portion of trajectories classified into the model $\mathcal M_j$ upon the true model $\mathcal M_i$.}
\label{fig:confusion}
\end{figure}

Figure~\ref{fig:confusion} shows the confusion matrices summarizing the performance of the Bayesian model classification upon the above-prepared data set. 
When $\mathrm{SNR}\geq 2$, our Bayesian inference gives high true positive rates for all three processes. Especially, the L\'evy walk trajectory is remarkably well classified by our Bayesian method. On the contrary, if the noise level is significantly high ($\mathrm{SNR}=1$), the L\'evy walk is the most poorly inferred model among the three models. In this case, the Gaussian noise tends to make the L\'evy walk falsely detected as FBM (a Gaussian and stationary-incremental process). The high true positive rates for FBM and SBM at $\mathrm{SNR}=1$, comparable to those at $\mathrm{SNR}=2$ and $10$, are attributed to the same effect.

For more quantitative analysis of the model classification, we study error rates of the L\'evy walk model as a function of $N_\mathrm{step}$, $\alpha$, and SNR. We estimate the false discovery rate~(FDR) and false negative rate~(FNR) from the confusion matrices. The former is the portion of FBM and SBM trajectories in the set of trajectories classified into LW, and the latter is the portion of misidentified L\'evy walk trajectories (as FBM or SBM) in the set of LW trajectories.

\begin{figure}
\centering
\includegraphics[width=1\textwidth]{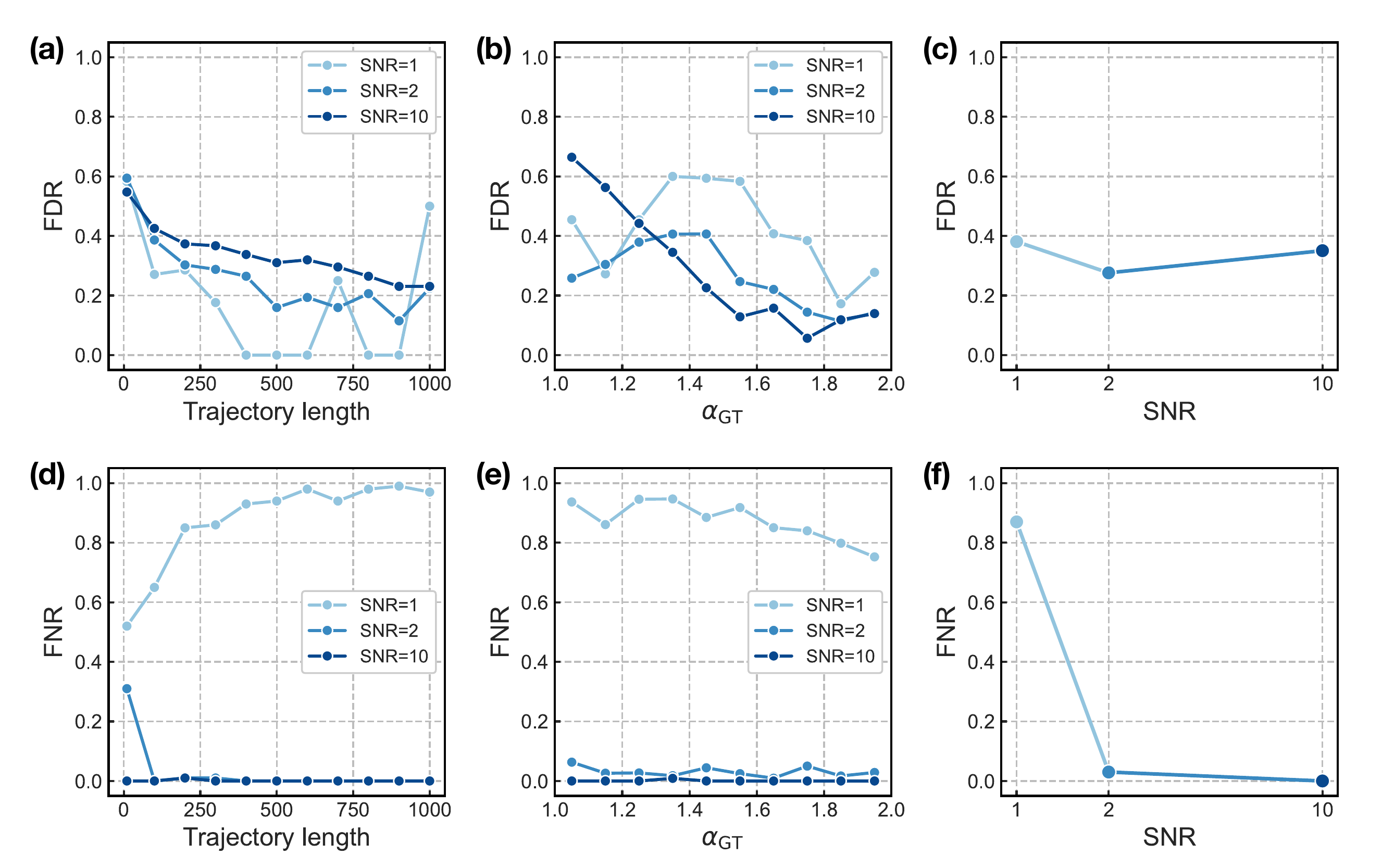}
\caption{Error rates obtained from confusion matrices as a function of the  trajectory length, $\alpha_\mathrm{GT}$, and SNR. (a) \& (d) False discovery rate and false negative rate of LWs as a function of the trajectory length. (b) \& (e) False discovery rate and false negative rate of LWs as a function of $\alpha_\mathrm{GT}$ with half bin width $|\Delta \alpha_\mathrm{GT}| = 0.05$. (c) \& (f) False discovery rate and false negative rate of LWs as a function of SNR. Each data point represents the averaged value over trajectory length and $\alpha_\mathrm{GT}$. }
\label{fig:errors}
\end{figure}

Fig.~\ref{fig:errors}(a) and (d), show the FDR and FNR as a function of trajectory length. The FDR tends to decrease with trajectory length irrespective of an SNR. Interestingly, FDR at $\mathrm{SNR}=1$ seems to be smaller than that for $\mathrm{SNR}\geq 2$ for most trajectory lengths. It appears that non-LW trajectories have an increased chance of being classified into our L\'evy walk model when an SNR is sufficiently large. This tendency can be seen in Fig.~\ref{fig:confusion} and Fig.~\ref{fig:errors}(c). The FNR is shown to be very small (except for $\mathrm{SNR}=1$) and tends to decrease with the trajectory length. We note that FNR behaves against the noise in the opposite way with FDR, such that a higher noise level leads to an increased FNR. As explained, at the highest noise level ($\mathrm{SNR}=1$), the LW trajectories are apt to be classified into FBM. 
Here, increasing trajectory length elevates the FNR because longer trajectories contain more evidence on the noise contribution and, subsequently, increase the chance of  being falsely detected as FBM.

In Fig.~\ref{fig:errors}(b) \& (d), we examine the variation of FDR and FNR as a function of $\alpha_\mathrm{GT}$. There is a clear tendency that the FDR increases as the ground true $\alpha$ approaches unity (Fig.~\ref{fig:errors}(b)). This is because when $\alpha=1$ the three models converge to the same process, i.e., Brownian motion. Hence, when a given (anomalous) diffusion process has $\alpha\approx1$, it is very difficult to pinpoint the correct underlying diffusion model. Ideally, the FDR should be $2/3$ at  $\alpha_\mathrm{GT}=1$ if there are three models. However, FNR (at $\mathrm{SNR}=2$ and $10$) is rarely affected by the ground truth anomaly exponent. We find from further analysis of the data that ambiguous Brownian-like trajectories around $\alpha_\mathrm{GT}\sim 1$ tends to be classified into LW than FBM or SBM. The rise of FNR at $\mathrm{SNR}=1$ is the falsely detected FBM from the LW trajectories.


\section{Discussion \& Summary} \label{sec:discussion}

In this work, we have proposed a hidden Markov model for L\'evy walks (LWs) and, using this model, developed Bayesian inference tools for the analysis of LW-like trajectory data. The essential part of our theory was to approximate a renewal process governed by a power-law sojourn time distribution with a Markovian decomposition method~\cite{goychuk2009}. We have shown that a power-law flight-time distribution can be generated by this method, enabling us to construct the likelihood functions of an LW. Using the Bayesian inference tool, we have demonstrated to extract the value of the power-law exponent of flight-time distribution ($\gamma=3-\alpha$) or, equivalently, the anomaly exponent $\alpha$ from given LW-like trajectories (containing background noises). It has been confirmed that the Bayesian inference method outperforms the traditional fitting method by extracting $\alpha$ from an MSD. When the noise level is negligible ($\mathrm{SNR}=10$), the accuracy of parameter estimation is as high as that by the top-performing machine learning methods recently developed~\cite{interactive, andi2021}. For the case of 2D L\'evy walk trajectories, remarkably, our Bayesian inference method is shown to perform better than the top-performing machine learning tool (which will be separately reported elsewhere). 
Apart from the parameter estimation, we have employed our Bayesian method to classify anomalous diffusion models. For three distinct diffusion processes (LW, FBM, and SBM) superimposed with noises, our Bayesian method successfully infers the correct model with a high true positive rate when the noise level is low ($\mathrm{SNR}=10,~2$).  



Our Bayesian inference method based on the hidden Markov model can be extended to analyzing multi-dimensional data or potentially applied to other similar problems. Below we provide some remarks on these issues. 

Firstly, it is straightforward to generalize the current model for the analysis of two-dimensional (2D) L\'evy walks. The 2D LWs are applicable to the modeling of cell motility on the surface~\cite{yang2011, li2015}, the projected 2D motion of actively transported cargoes in the cell~\cite{fedotov2018, chen2015}, swarming dynamics of bacteria~\cite{ariel2015, be2019}, and various foraging dynamics in ecology~\cite{raichlen2014, reynolds2009, garg2021, humphries2010}. We can define a L\'evy walk in 2D and construct a hidden Markov model for 2D LWs. The method is briefly sketched in the \ref{sec:2dlevy}. A full investigation of the Bayesian inference method to 2D LWs will be reported elsewhere. One remarkable result is that the estimation of $\alpha$ is more accurate in 2D than in 1D. This is attributed to the fact that successive flight events are more easily identified in 2D, which leads to the increased performance in the parameter estimation of 2D LWs at a higher SNR.

Secondly, our L\'evy walk Bayesian model can be easily expanded to include composite correlated random walk (CCRW) processes via an analogous Markovian decomposite method introduced in this work. The CCRW is a multi-state random walk in which each state has a short-term (Markovian) directional memory leading to an exponential step-length distribution~\cite{auger2015, codling2008}. The foraging dynamics of animals is an important example of the CCRW~\cite{jager2011, Jansen2012, reynolds2014}. The stochastic properties of a CCRW are governed by the superposition rule of the multiple Markovian states. In the framework of CCRWs, the L\'evy walk can be regarded as a special model where the superposition is specified by Eqs.~(\ref{eq:rate}) and (\ref{eq:transition}), leading to a power-law flight-time (or step-length) distribution. The CCRW may have a different step-length distribution depending on how multiple Markovian states are superposed. By properly modeling this part, a specific CCRW can be constructed and inferred by the corresponding Bayesian method. A potential application is to identify a correct diffusion model for animal's foraging dynamics. There is a stimulating effort in ecology for modeling the foraging motion~\cite{jager2011, Jansen2012, reynolds2014, andy2014, simon2006, simon2007}. Often the analysis of the data was based on the step-length distribution (or the inverse cumulative distribution of the step lengths), which led to a debate on whether the data follows a L\'evy walk having a power-law step length distribution or a different CCRW~\cite{Jansen2012, auger2015, jager2011, reynolds2014}. Using the Bayesian method, one can analyze the data in a different manner and efficiently differentiate between several candidate models.

Thirdly, our hidden Markov model is applicable to construct Bayesian inference methods for variant L\'evy walk models. For example, one can expand the current hidden Markov model for a L\'evy walk with fluctuating velocities or a L\'evy walk with a rest~\cite{zaburdaev2015}. The latter model was shown to explain the fluid dynamics in a rotating plate and the transport of mRNA-protein complex particles in a neuron cell~\cite{song2018}. The Bayesian method will be helpful for extracting the information of parameters in the flight-time and rest-time distributions. Beyond the L\'evy walk, once the noise correlation in the successive displacement is adequately incorporated, the current hidden Markov model can be applied to continuous-time random walks with a power-law waiting time distribution.

Finally we comment on some potential technical issues of the current Bayesian approach. The high computational cost is often required for Bayesian inference, which needs to sample parameters from high dimensional posterior distributions and calculate likelihood functions for every sampled parameter. It costs hundreds to thousands of seconds to quantify a trajectory of hundreds of steps. Notwithstanding, it is computationally feasible to handle thousands of trajectories of which step length is of $<1000$. In our study, we analyzed such trajectory data within a day using a 40-multicore workstation. Additionally, a relevant task is to improve the Bayesian inference method under a strong noise signal. In this work, we have found that the Bayesian inference analysis is vulnerable to noise when $\mathrm{SNR}\sim 1$. We anticipate that if the noise effect term in the likelihood function [Eq.~(\ref{eq:emission})] was incorporated without approximation, the Bayesian inference would have a better performance to noisy trajectories under the condition of strong noise signals. We leave this task as future work.

\section*{Acknowledgements}
This work was supported by the National Research Foundation (NRF) of Korea (No.~2020R1A2C4002490).

\newpage

\section*{Appendix}
\appendix
\section{Continuous-time Markov chain approximation of continuous target distribution} \label{sec:powerlaw_deriv}

Here we briefly show the mathematical derivation of Eq.~(\ref{eq:approxtarget}). The continuous target distribution is given by the inter-arrival time distributions and the weights as
\begin{equation}\label{eqA1}
\begin{aligned}
    \psi^\mathrm{target}_\mathrm{c}(\tau; \alpha) & \approx  \sum_{i=1}^{N_k} P_{i; \alpha} \phi_i (\tau)=
\sum_{i=1}^{N_k} \left( \frac{k_i^{3-\alpha}e^{-k_i}}{\sum_{j=1}^{N_k} k_j^{3-\alpha}e^{-k_j}} \right) k_i e^{-k_i(\tau-1)} \\
& =  \frac{1}{\sum_{j=1}^{N_k} k_j^{3-\alpha} e^{-k_j} } \sum_{i=1}^{N_k} {k_i^{4-\alpha} e^{-k_i\tau} }.
\end{aligned}
\end{equation}
In this expression, the denominator $\sum_{j=1}^{N_k} k_{j}^{3-\alpha} e^{-k_j}$ is estimated under the assumption of a sufficiently large $N_k$ to
\begin{equation}
\begin{aligned}
    \sum_{j=1}^{N_k}  \left(\frac{k_{\mathrm{fast}}}{b^j} \right) ^{3-\alpha} e^{- \frac{k_{\mathrm{fast}} }{b^j}} & \approx \int_{\frac{1}{2}} ^{\infty}dx \left(\frac{k_{\mathrm{fast}}}{b^x} \right) ^{3-\alpha} e^{- \frac{k_{\mathrm{fast}} }{b^x}}  \\
    & = \frac{1}{\ln b} \gamma \left( 3-\alpha , k_{\mathrm{fast} } / \sqrt{b} \right)
\end{aligned}
\end{equation}
where $\gamma(s,x)=\int_0^x t^{s-1}e^{-t} dt$ is the lower incomplete gamma function. The numerator can be calculated in the same way. This approximation yields
\begin{equation}
\begin{aligned}
        \sum_{i=1}^{N_k}P_{i;\alpha}\phi_i(\tau) \approx \frac{\gamma \left( 4-\alpha, k_{\mathrm{fast}}\tau /\sqrt{b} \right)}{\gamma\left(3-\alpha, k_{\mathrm{fast}}/\sqrt{b}\right)} \frac{1}{\tau^{4-\alpha}}.
\end{aligned}
\end{equation}
When $k_\mathrm{fast}$ becomes considerably large, the prefactor approaches to
\begin{equation}
\frac{\gamma \left(4-\alpha,k_{\mathrm{fast}} \tau/\sqrt{b}  \right)}{\gamma \left(3-\alpha,k_{\mathrm{fast}}/\sqrt{b}  \right)} \approx \frac{\Gamma(4-\alpha)}{\Gamma(3-\alpha)} = 3-\alpha.
\end{equation}
Thus, $\sum_{i=1}^{N_k} P_{i;\alpha} \phi_i(\tau)$ becomes the normalized continuous target distribution given by Eq.~(\ref{eq:exactcont}).

\section{Modeling two-dimensional L\'evy walks} \label{sec:2dlevy}

\begin{figure}
\centering
\includegraphics[width=1\textwidth]{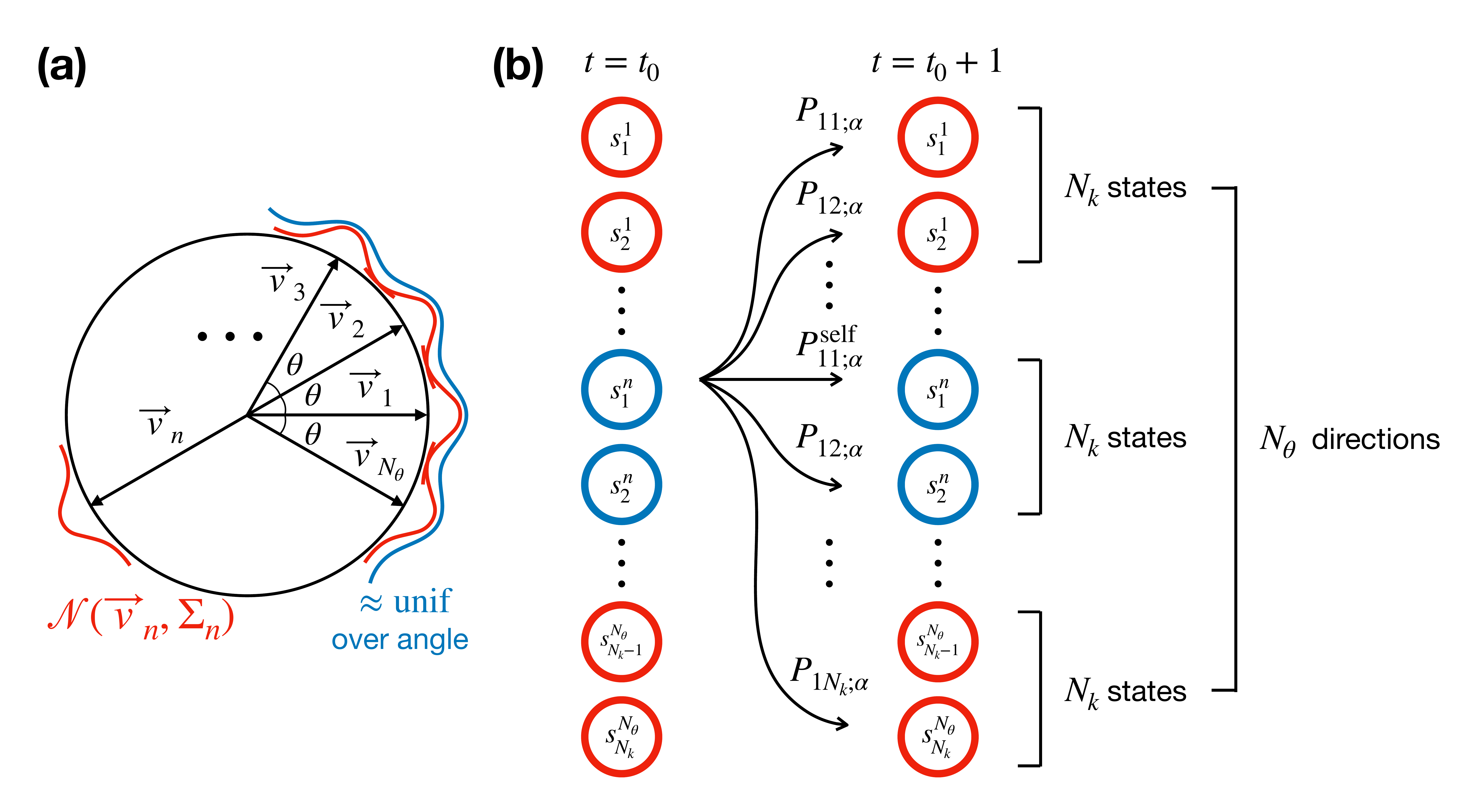}
\caption{(a) The velocity states in our 2D L\'evy walk. There are $N_\theta$ discrete velocity states specified by $\vec{v}_n$ ($n=1,2,\ldots,N_\theta$).
We allow the orientation of $\vec{v}_n$ to fluctuate around the unit vector $(\cos((n-1)\theta),\sin((n-1)\theta))$, which is modeled by a Gaussian fluctuation $\mathcal N (\vec{v}_n, \Sigma_n)$. The superposition of the Gaussian fluctuations (the blue solid line) approximates the uniformly distributed velocity states over angle. (b) The Markov chain model for the 2D discrete-time L\'evy walk. Each velocity states in (a) have the latent state $s^n_i$ with $i=1,2,\ldots,N_k$. The superscript $n$ denotes the velocity state. The transition probability from the state $s_i^m$ to $s_j^n$ is given by $P_{ij;\alpha}$ [Eq.~(\ref{eq:2Dtransprob})], and the self-transition probability is given by $P_{ii;\alpha}^\mathrm{self}$ [Eq.~(\ref{eq:2Dselftransprob})].}
\label{fig:2Dscheme}
\end{figure}

In the Markov chain approximation of 1D L\'evy walk in the main text, we need two independent latent state sets ($s^+_i$ and $s^-_i$ with $i=1,2,\ldots,N_k$) for the flight of positive and negative directions, respectively. In the 2D L\'evy walk~\cite{zaburdaev2016}, we set there are a total of $N_\theta$ velocity orientations (see Fig.~\ref{fig:2Dscheme}(a)).
A random walker can choose a random direction from the velocity vectors $\{\vec{v_1}, \vec{v_2}, \cdots, \vec{v_{N_\theta}}\}$ with a probability of $1/N_\theta$. We allow a velocity vector $\vec{v_n}$ to have uncertainty in its direction, which is modeled by a Gaussian fluctuation (see the red lines in Fig.~\ref{fig:2Dscheme}(a)). We obtain a nearly uniform distribution over the velocity direction~(blue line in Fig.~\ref{fig:2Dscheme}(a)) once all the Gaussian fluctuations $\mathcal{N}(\vec{v}_n, \Sigma_n)$ are superposed. 

There are a total of $N_\theta$ number of velocity states in the 2D L\'evy walk, where there are $N_k$ latent states with independent transition rates $k_i$ $(i=1,2,\ldots,N_k)$ for each velocity state. The probability of self-transition is
\begin{equation}\label{eq:2Dselftransprob}
    P_{ii;\alpha}^{\mathrm{self}} = e^{-k_i} + \frac{(1-e^{-k_i}) \cdot P_{i;\alpha}}{N_\theta},
\end{equation}
and the transition probability to other states reads
\begin{equation}\label{eq:2Dtransprob}
    P_{ij;\alpha} = \frac{(1-e^{-k_i}) \cdot P_{j;\alpha}}{N_\theta}.
\end{equation}
\begin{figure}
\centering
\includegraphics[width=1\textwidth]{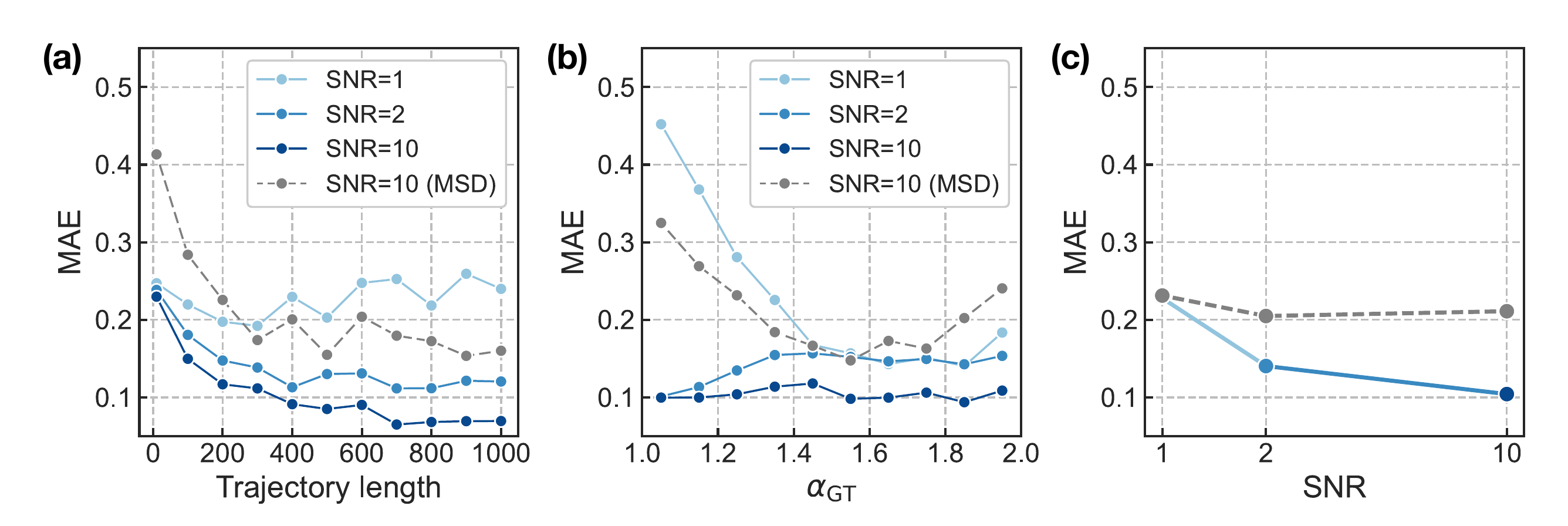}
\caption{Parameter estimation for 2D L\'evy walk trajectories. (a) Mean absolute error (MAE) as a function of the trajectory length. (b) MAE as a function of the ground-truth anomaly exponent $\alpha_{\mathrm{GT}}$. (c) MAE as a function of SNR. The dotted line is the result from the Bayesian inference and the dot-dashed line from the fitting of MSD curves.}
\label{fig:2dAE}
\end{figure}
The emission probability of an observation $\Delta \vec{x}_t = \vec{x}_{t+1} - \vec{x}_t$ from the velocity state $\vec{v}_n$ is given by
\begin{equation}\label{eq:2Demission}
    P(\Delta \vec{x}_t | \mathcal{M}_\mathrm{L\acute{e}vy}, \vec{v}_n, \Sigma_n) = \frac{\exp \left ( -\frac{1}{2} (\Delta \vec{x}_t - \vec{v}_n)^\mathrm{T} \Sigma_n^{-1} (\Delta \vec{x}_t - \vec{v}_n) \right )}{\sqrt{(2\pi)^d |\Sigma_n|}},
\end{equation}
where $\Sigma_n = R(\theta_n) \Sigma_1 R(\theta_n)^\mathrm{T}$ is a covariance matrix for Gaussian fluctuations of $\vec{v}_n$, $\theta_n$ is an angle from $x$-axis of a vector $\vec{v}_n$, and $R(\theta_n)$ is a rotation matrix. From the Markov chain illustrated in Fig.~\ref{fig:2Dscheme}(b) with transition probabilities (\ref{eq:2Dselftransprob}) and (\ref{eq:2Dtransprob}) and the emission probability (\ref{eq:2Demission}), the likelihood function for 2D L\'evy walks can be calculated using forward algorithm as in Eqs.~(\ref{eq:seqL})-(\ref{eq:likelihood}). The likelihood function for a 3D L\'evy walk model can also be obtained in a similar way.

Figure~\ref{fig:2dAE} shows the performance of the parameter estimation. While most of the results are similar to the one-dimensional case, we notice some interesting features: The accuracy of parameter estimation has been improved at $\mathrm{SNR}=10$ compared to the corresponding case in 1D. Figure~\ref{fig:2dAE} shows that mean absolute error decays faster than 1D as trajectory length increases. Moreover, the MAE does not change upon the variation of anomaly exponents (Fig.~\ref{fig:2dAE}(b)). At $\mathrm{SNR}=10$, the overall MAE decreases by $\sim0.02$ compared to the 1D case 
(Fig.~\ref{fig:2dAE}(c)).

\bibliographystyle{unsrt} 
\bibliography{reference}

\begin{thebibliography}{10}

\bibitem{metzler2014}
Ralf Metzler, Jae-Hyung Jeon, Andrey~G. Cherstvy, and Eli Barkai.
\newblock Anomalous diffusion models and their properties: non-stationarity{,}
  non-ergodicity{,} and ageing at the centenary of single particle tracking.
\newblock {\em Phys. Chem. Chem. Phys.}, 16:24128--24164, 2014.

\bibitem{saxton1997}
Michael~J Saxton and Ken Jacobson.
\newblock Single-particle tracking: applications to membrane dynamics.
\newblock {\em Annual review of biophysics and biomolecular structure},
  26(1):373--399, 1997.

\bibitem{golding2006}
Ido Golding and Edward~C. Cox.
\newblock Physical nature of bacterial cytoplasm.
\newblock {\em Phys. Rev. Lett.}, 96:098102, Mar 2006.

\bibitem{jeon2011}
Jae-Hyung Jeon, Vincent Tejedor, Stas Burov, Eli Barkai, Christine
  Selhuber-Unkel, Kirstine Berg-S\o{}rensen, Lene Oddershede, and Ralf Metzler.
\newblock In vivo anomalous diffusion and weak ergodicity breaking of lipid
  granules.
\newblock {\em Phys. Rev. Lett.}, 106:048103, Jan 2011.

\bibitem{rienzo2014}
Carmine Di~Rienzo, Vincenzo Piazza, Enrico Gratton, Fabio Beltram, and
  Francesco Cardarelli.
\newblock Probing short-range protein brownian motion in the cytoplasm of
  living cells.
\newblock {\em Nature communications}, 5(1):1--8, 2014.

\bibitem{krapf2019}
Diego Krapf and Ralf Metzler.
\newblock Strange interfacial molecular dynamics.
\newblock {\em Physics today}, 72(9):48--55, 2019.

\bibitem{lee2021}
Daniel~SW Lee, Ned~S Wingreen, and Clifford~P Brangwynne.
\newblock Chromatin mechanics dictates subdiffusion and coarsening dynamics of
  embedded condensates.
\newblock {\em Nature Physics}, 17(4):531--538, 2021.

\bibitem{park2021}
Seongyu Park, O-chul Lee, Xavier Durang, and Jae-Hyung Jeon.
\newblock A mini-review of the diffusion dynamics of dna-binding proteins:
  experiments and models.
\newblock {\em Journal of the Korean Physical Society}, 78(5):408--426, 2021.

\bibitem{wang2012}
Bo~Wang, James Kuo, Sung~Chul Bae, and Steve Granick.
\newblock When brownian diffusion is not gaussian.
\newblock {\em Nature Materials}, 11(6):481--485, 2012.

\bibitem{cherstvy2019}
Andrey~G. Cherstvy, Samudrajit Thapa, Caroline~E. Wagner, and Ralf Metzler.
\newblock Non-gaussian{,} non-ergodic{,} and non-fickian diffusion of tracers
  in mucin hydrogels.
\newblock {\em Soft Matter}, 15:2526--2551, 2019.

\bibitem{kim2020}
Won~Kyu Kim, Richard Chudoba, Sebastian Milster, Rafael Roa, Matej
  Kandu{\v{c}}, and Joachim Dzubiella.
\newblock Tuning the selective permeability of polydisperse polymer networks.
\newblock {\em Soft Matter}, 16(35):8144--8154, 2020.

\bibitem{ariel2015}
Gil Ariel, Amit Rabani, Sivan Benisty, Jonathan~D Partridge, Rasika~M Harshey,
  and Avraham Be'Er.
\newblock Swarming bacteria migrate by l{\'e}vy walk.
\newblock {\em Nature communications}, 6(1):1--6, 2015.

\bibitem{ipina2019}
Emiliano~Perez Ipi{\~n}a, Stefan Otte, Rodolphe Pontier-Bres, Dorota Czerucka,
  and Fernando Peruani.
\newblock Bacteria display optimal transport near surfaces.
\newblock {\em Nature Physics}, 15(6):610--615, 2019.

\bibitem{patteson2015}
AE~Patteson, Arvind Gopinath, M~Goulian, and PE~Arratia.
\newblock Running and tumbling with e. coli in polymeric solutions.
\newblock {\em Scientific reports}, 5(1):1--11, 2015.

\bibitem{humphries2010}
Nicolas~E Humphries, Nuno Queiroz, Jennifer~RM Dyer, Nicolas~G Pade, Michael~K
  Musyl, Kurt~M Schaefer, Daniel~W Fuller, Juerg~M Brunnschweiler, Thomas~K
  Doyle, Jonathan~DR Houghton, et~al.
\newblock Environmental context explains l{\'e}vy and brownian movement
  patterns of marine predators.
\newblock {\em Nature}, 465(7301):1066--1069, 2010.

\bibitem{jager2011}
Monique de~Jager, Franz~J. Weissing, Peter M.~J. Herman, Bart~A. Nolet, and
  Johan van~de Koppel.
\newblock L{\'e}vy walks evolve through interaction between movement and
  environmental complexity.
\newblock {\em Science}, 332(6037):1551--1553, 2011.

\bibitem{wergen2011}
Gregor Wergen, Miro Bogner, and Joachim Krug.
\newblock Record statistics for biased random walks, with an application to
  financial data.
\newblock {\em Phys. Rev. E}, 83:051109, May 2011.

\bibitem{plerou2000}
Vasiliki Plerou, Parameswaran Gopikrishnan, Lu\'{\i}s~A. Nunes~Amaral, Xavier
  Gabaix, and H.~Eugene~Stanley.
\newblock Economic fluctuations and anomalous diffusion.
\newblock {\em Phys. Rev. E}, 62:R3023--R3026, Sep 2000.

\bibitem{brown1828}
Robert Brown F.R.S. Hon. M.R.S.E. \& R.I.~Acad. V.P.L.S.
\newblock Xxvii. a brief account of microscopical observations made in the
  months of june, july and august 1827, on the particles contained in the
  pollen of plants; and on the general existence of active molecules in organic
  and inorganic bodies.
\newblock {\em The Philosophical Magazine}, 4(21):161--173, 1828.

\bibitem{einstein1956}
Albert Einstein.
\newblock {\em Investigations on the Theory of the Brownian Movement}.
\newblock Courier Corporation, 1956.

\bibitem{smoluchowski1906}
M.~von Smoluchowski.
\newblock Zur kinetischen theorie der brownschen molekularbewegung und der
  suspensionen.
\newblock {\em Annalen der Physik}, 326(14):756--780, 1906.

\bibitem{reverey2015}
Julia~F. Reverey, Jae-Hyung Jeon, Han Bao, Matthias Leippe, Ralf Metzler, and
  Christine Selhuber-Unkel.
\newblock Superdiffusion dominates intracellular particle motion in the
  supercrowded cytoplasm of pathogenic acanthamoeba castellanii.
\newblock {\em Scientific Reports}, 5(1):11690, 2015.

\bibitem{lampo2017}
Thomas~J. Lampo, Stella Stylianidou, Mikael~P. Backlund, Paul~A. Wiggins, and
  Andrew~J. Spakowitz.
\newblock Cytoplasmic rna-protein particles exhibit non-gaussian subdiffusive
  behavior.
\newblock {\em Biophysical Journal}, 112(3):532--542, 2017.

\bibitem{scher1975}
Harvey Scher and Elliott~W. Montroll.
\newblock Anomalous transit-time dispersion in amorphous solids.
\newblock {\em Phys. Rev. B}, 12:2455--2477, Sep 1975.

\bibitem{montroll1969}
Elliott~W. Montroll.
\newblock Random walks on lattices. iii. calculation of first‐passage times
  with application to exciton trapping on photosynthetic units.
\newblock {\em Journal of Mathematical Physics}, 10(4):753--765, 1969.

\bibitem{mandelbrot1968}
Benoit~B Mandelbrot and John~W Van~Ness.
\newblock Fractional brownian motions, fractional noises and applications.
\newblock {\em SIAM review}, 10(4):422--437, 1968.

\bibitem{zaburdaev2015}
Vasily Zaburdaev, S~Denisov, and J~Klafter.
\newblock L{\'e}vy walks.
\newblock {\em Reviews of Modern Physics}, 87(2):483, 2015.

\bibitem{massignan2014}
P.~Massignan, C.~Manzo, J.~A. Torreno-Pina, M.~F. Garc\'{\i}a-Parajo,
  M.~Lewenstein, and G.~J. Lapeyre.
\newblock Nonergodic subdiffusion from brownian motion in an inhomogeneous
  medium.
\newblock {\em Phys. Rev. Lett.}, 112:150603, Apr 2014.

\bibitem{lim2002}
SC~Lim and SV~Muniandy.
\newblock Self-similar gaussian processes for modeling anomalous diffusion.
\newblock {\em Physical Review E}, 66(2):021114, 2002.

\bibitem{jeon2014}
Jae-Hyung Jeon, Aleksei~V Chechkin, and Ralf Metzler.
\newblock Scaled brownian motion: a paradoxical process with a time dependent
  diffusivity for the description of anomalous diffusion.
\newblock {\em Physical Chemistry Chemical Physics}, 16(30):15811--15817, 2014.

\bibitem{metzler2009analysis}
Ralf Metzler, Vincent Tejedor, J-H Jeon, Y~He, WH~Deng, S~Burov, and E~Barkai.
\newblock Analysis of single particle trajectories: From normal to anomalous
  diffusion.
\newblock {\em Acta Physica Polonica B}, 40(5), 2009.

\bibitem{Jeon2013}
Jae-Hyung Jeon, Natascha Leijnse, Lene~B Oddershede, and Ralf Metzler.
\newblock Anomalous diffusion and power-law relaxation of the time averaged
  mean squared displacement in worm-like micellar solutions.
\newblock {\em New Journal of Physics}, 15(4):045011, apr 2013.

\bibitem{safdari2015}
Hadiseh Safdari, Andrey~G Cherstvy, Aleksei~V Chechkin, Felix Thiel, Igor~M
  Sokolov, and Ralf Metzler.
\newblock Quantifying the non-ergodicity of scaled brownian motion.
\newblock {\em Journal of Physics A: Mathematical and Theoretical},
  48(37):375002, aug 2015.

\bibitem{cherstvy2015}
Andrey~G. Cherstvy and Ralf Metzler.
\newblock Ergodicity breaking and particle spreading in noisy heterogeneous
  diffusion processes.
\newblock {\em The Journal of Chemical Physics}, 142(14):144105, 2015.

\bibitem{burov2011}
Stas Burov, Jae-Hyung Jeon, Ralf Metzler, and Eli Barkai.
\newblock Single particle tracking in systems showing anomalous diffusion: the
  role of weak ergodicity breaking.
\newblock {\em Phys. Chem. Chem. Phys.}, 13:1800--1812, 2011.

\bibitem{ernst2014}
Dominique Ernst, Jürgen Köhler, and Matthias Weiss.
\newblock Probing the type of anomalous diffusion with single-particle
  tracking.
\newblock {\em Phys. Chem. Chem. Phys.}, 16:7686--7691, 2014.

\bibitem{kepten2015}
E.~Kepten, A.~Weron, G.~Sikora, K.~Burnecki, and Y.~Garini.
\newblock Guidelines for the fitting of anomalous diffusion mean square
  displacement graphs from single particle tracking experiments.
\newblock {\em PLoS One}, 10(2):e0117722, 2015.

\bibitem{jamali2021}
Vida Jamali, Cory Hargus, Assaf Ben-Moshe, Amirali Aghazadeh, Hyun~Dong Ha,
  Kranthi~K. Mandadapu, and A.~Paul Alivisatos.
\newblock Anomalous nanoparticle surface diffusion in lctem is revealed by deep
  learning-assisted analysis.
\newblock {\em Proceedings of the National Academy of Sciences}, 118(10), 2021.

\bibitem{gorka2020}
Gorka Mu{\~{n}}oz-Gil, Miguel~Angel Garcia-March, Carlo Manzo, Jos{\'{e}}~D
  Mart{\'{\i}}n-Guerrero, and Maciej Lewenstein.
\newblock Single trajectory characterization via machine learning.
\newblock {\em New Journal of Physics}, 22(1):013010, jan 2020.

\bibitem{cichos2020}
Frank Cichos, Kristian Gustavsson, Bernhard Mehlig, and Giovanni Volpe.
\newblock Machine learning for active matter.
\newblock {\em Nature Machine Intelligence}, 2(2):94--103, 2020.

\bibitem{granik2019}
Naor Granik, Lucien~E. Weiss, Elias Nehme, Maayan Levin, Michael Chein, Eran
  Perlson, Yael Roichman, and Yoav Shechtman.
\newblock Single-particle diffusion characterization by deep learning.
\newblock {\em Biophysical Journal}, 117(2):185--192, 2019.

\bibitem{martin2016}
Martin Lysy, Natesh~S. Pillai, David~B. Hill, M.~Gregory Forest, John W.~R.
  Mellnik, Paula~A. Vasquez, and Scott~A. McKinley.
\newblock Model comparison and assessment for single particle tracking in
  biological fluids.
\newblock {\em Journal of the American Statistical Association},
  111(516):1413--1426, 2016.

\bibitem{thapa2018}
Samudrajit Thapa, Michael~A. Lomholt, Jens Krog, Andrey~G. Cherstvy, and Ralf
  Metzler.
\newblock Bayesian analysis of single-particle tracking data using the
  nested-sampling algorithm: maximum-likelihood model selection applied to
  stochastic-diffusivity data.
\newblock {\em Phys. Chem. Chem. Phys.}, 20:29018--29037, 2018.

\bibitem{krog2018}
Jens Krog, Lars~H Jacobsen, Frederik~W Lund, Daniel Wüstner, and Michael~A
  Lomholt.
\newblock Bayesian model selection with fractional brownian motion.
\newblock {\em Journal of Statistical Mechanics: Theory and Experiment},
  2018(9):093501, sep 2018.

\bibitem{krog2017}
Jens Krog and Michael~A. Lomholt.
\newblock Bayesian inference with information content model check for langevin
  equations.
\newblock {\em Phys. Rev. E}, 96:062106, Dec 2017.

\bibitem{auger2015}
Marie Auger-Méthé, Andrew~E. Derocher, Michael~J. Plank, Edward~A. Codling,
  and Mark~A. Lewis.
\newblock Differentiating the lévy walk from a composite correlated random
  walk.
\newblock {\em Methods in Ecology and Evolution}, 6(10):1179--1189, 2015.

\bibitem{raichlen2014}
David~A. Raichlen, Brian~M. Wood, Adam~D. Gordon, Audax Z.~P. Mabulla, Frank~W.
  Marlowe, and Herman Pontzer.
\newblock Evidence of l{\'e}vy walk foraging patterns in human
  hunter{\textendash}gatherers.
\newblock {\em Proceedings of the National Academy of Sciences},
  111(2):728--733, 2014.

\bibitem{reynolds2009}
A.~M. Reynolds and C.~J. Rhodes.
\newblock The lévy flight paradigm: random search patterns and mechanisms.
\newblock {\em Ecology}, 90(4):877--887, 2009.

\bibitem{garg2021}
Ketika Garg and Christopher~T. Kello.
\newblock Efficient lévy walks in virtual human foraging.
\newblock {\em Scientific Reports}, 11(1):5242, 2021.

\bibitem{wosniack2017}
Marina~E. Wosniack, Marcos~C. Santos, Ernesto~P. Raposo, Gandhi~M. Viswanathan,
  and Marcos G.~E. da~Luz.
\newblock The evolutionary origins of lévy walk foraging.
\newblock {\em PLOS Computational Biology}, 13(10):1--31, 10 2017.

\bibitem{rhee2011}
Injong Rhee, Minsu Shin, Seongik Hong, Kyunghan Lee, Seong~Joon Kim, and Song
  Chong.
\newblock On the levy-walk nature of human mobility.
\newblock {\em IEEE/ACM Transactions on Networking}, 19(3):630--643, 2011.

\bibitem{focardi2009}
Stefano Focardi, Paolo Montanaro, and Elena Pecchioli.
\newblock Adaptive lévy walks in foraging fallow deer.
\newblock {\em PLOS ONE}, 4(8):1--6, 08 2009.

\bibitem{huo2021}
Haiyan Huo, Rui He, Rongjing Zhang, Junhua Yuan, and Gladys Alexandre.
\newblock Swimming escherichia coli cells explore the environment by l\'evy
  walk.
\newblock {\em Applied and Environmental Microbiology}, 87(6):e02429--20, 2021.

\bibitem{song2018}
Minho~S. Song, Hyungseok~C. Moon, Jae-Hyung Jeon, and Hye~Yoon Park.
\newblock Neuronal messenger ribonucleoprotein transport follows an aging lévy
  walk.
\newblock {\em Nature Communications}, 9(1):344, 2018.

\bibitem{gal2010}
Naama Gal and Daphne Weihs.
\newblock Experimental evidence of strong anomalous diffusion in living cells.
\newblock {\em Phys. Rev. E}, 81:020903, Feb 2010.

\bibitem{chen2015}
Kejia Chen, Bo~Wang, and Steve Granick.
\newblock Memoryless self-reinforcing directionality in endosomal active
  transport within living cells.
\newblock {\em Nature Materials}, 14(6):589--593, 2015.

\bibitem{stefani2009}
Fernando~D. Stefani, Jacob~P. Hoogenboom, and Eli Barkai.
\newblock Beyond quantum jumps: Blinking nanoscale light emitters.
\newblock {\em Physics Today}, 62(2):34--39, 2009.

\bibitem{barthelemy2008}
Pierre Barthelemy, Jacopo Bertolotti, and Diederik~S. Wiersma.
\newblock A lévy flight for light.
\newblock {\em Nature}, 453(7194):495--498, 2008.

\bibitem{solomon1993}
T.~H. Solomon, Eric~R. Weeks, and Harry~L. Swinney.
\newblock Observation of anomalous diffusion and l\'evy flights in a
  two-dimensional rotating flow.
\newblock {\em Phys. Rev. Lett.}, 71:3975--3978, Dec 1993.

\bibitem{klafter1994}
J.~Klafter and G.~Zumofen.
\newblock L\'evy statistics in a hamiltonian system.
\newblock {\em Phys. Rev. E}, 49:4873--4877, Jun 1994.

\bibitem{geisel1985}
T.~Geisel, J.~Nierwetberg, and A.~Zacherl.
\newblock Accelerated diffusion in josephson junctions and related chaotic
  systems.
\newblock {\em Phys. Rev. Lett.}, 54:616--619, Feb 1985.

\bibitem{gelman2013}
Andrew Gelman, John~B Carlin, Hal~S Stern, David~B Dunson, Aki Vehtari, and
  Donald~B Rubin.
\newblock {\em Bayesian data analysis}.
\newblock CRC press, 2013.

\bibitem{Bayes1763}
Thomas Bayes.
\newblock An essay towards solving a problem in the doctrine of chances.
\newblock {\em Philosophical Transactions of the Royal Society of London},
  53:370--418, 1763.

\bibitem{neal2001}
Radford~M. Neal.
\newblock Annealed importance sampling.
\newblock {\em Statistics and Computing}, 11(2):125--139, 2001.

\bibitem{BIT}
Bayesian inference for andi challenge.
\newblock \verb"https://github.com/mlomholt/andi".
\newblock accessed: 2021-06-03.

\bibitem{andipackage}
Gorka Muñoz-Gil, Borja Requena, Giovanni Volpe, Miguel~Angel Garcia-March, and
  Carlo Manzo.
\newblock {AnDiChallenge/ANDI\_datasets: Challenge 2020 release}, May 2021.

\bibitem{andi_expl}
Gorka Muñoz-Gil, Giovanni Volpe, Miguel~Angel García-March, Ralf Metzler,
  Maciej Lewenstein, and Carlo Manzo.
\newblock The anomalous diffusion challenge: single trajectory characterisation
  as a competition.
\newblock {\em Emerging Topics in Artificial Intelligence 2020}, Aug 2020.

\bibitem{goychuk2009}
Igor Goychuk.
\newblock Viscoelastic subdiffusion: From anomalous to normal.
\newblock {\em Phys. Rev. E}, 80:046125, Oct 2009.

\bibitem{Rabiner1989}
L.R. Rabiner.
\newblock A tutorial on hidden markov models and selected applications in
  speech recognition.
\newblock {\em Proceedings of the IEEE}, 77(2):257--286, 1989.

\bibitem{andi2021}
Gorka Muñoz-Gil, Giovanni Volpe, Miguel~Angel Garcia-March, Erez Aghion, Aykut
  Argun, Chang~Beom Hong, Tom Bland, Stefano Bo, J.~Alberto Conejero, Nicolás
  Firbas, Òscar Garibo~i Orts, Alessia Gentili, Zihan Huang, Jae-Hyung Jeon,
  Hélène Kabbech, Yeongjin Kim, Patrycja Kowalek, Diego Krapf, Hanna
  Loch-Olszewska, Michael~A. Lomholt, Jean-Baptiste Masson, Philipp~G. Meyer,
  Seongyu Park, Borja Requena, Ihor Smal, Taegeun Song, Janusz Szwabiński,
  Samudrajit Thapa, Hippolyte Verdier, Giorgio Volpe, Arthur Widera, Maciej
  Lewenstein, Ralf Metzler, and Carlo Manzo.
\newblock Objective comparison of methods to decode anomalous diffusion, 2021.
\newblock arXiv:2105.06766.

\bibitem{samuSBM}
Samudrajit Thapa.
\newblock Bayesian inference with scaled brownian motion.
\newblock {\em Journal of Physics A: Mathematical and Theoretical}, 2021.

\bibitem{interactive}
Andi interactive tool.
\newblock \verb"https://andi-app.herokuapp.com/andi_app".
\newblock accessed: 2021-06-03.

\bibitem{yang2011}
Taeseok~Daniel Yang, Jin-Sung Park, Youngwoon Choi, Wonshik Choi, Tae-Wook Ko,
  and Kyoung~J Lee.
\newblock Zigzag turning preference of freely crawling cells.
\newblock {\em PLoS One}, 6(6):e20255, 2011.

\bibitem{li2015}
Hui Li, Shuhong Qi, Honglin Jin, Zhongyang Qi, Zhihong Zhang, Ling Fu, and
  Qingming Luo.
\newblock Zigzag generalized levy walk: the in vivo search strategy of
  immunocytes.
\newblock {\em Theranostics}, 5(11):1275, 2015.

\bibitem{fedotov2018}
Sergei Fedotov, Nickolay Korabel, Thomas~A. Waigh, Daniel Han, and Victoria~J.
  Allan.
\newblock Memory effects and l\'evy walk dynamics in intracellular transport of
  cargoes.
\newblock {\em Phys. Rev. E}, 98:042136, Oct 2018.

\bibitem{be2019}
Avraham Be’er and Gil Ariel.
\newblock A statistical physics view of swarming bacteria.
\newblock {\em Movement ecology}, 7(1):1--17, 2019.

\bibitem{codling2008}
Edward~A Codling, Michael~J Plank, and Simon Benhamou.
\newblock Random walk models in biology.
\newblock {\em Journal of The Royal Society Interface}, 5(25):813--834, 2008.

\bibitem{Jansen2012}
Vincent A.~A. Jansen, Alla Mashanova, and Sergei Petrovskii.
\newblock Comment on {\textquotedblleft}l{\'e}vy walks evolve through
  interaction between movement and environmental
  complexity{\textquotedblright}.
\newblock {\em Science}, 335(6071):918--918, 2012.

\bibitem{reynolds2014}
Andy~M. Reynolds.
\newblock Mussels realize weierstrassian lévy walks as composite correlated
  random walks.
\newblock {\em Scientific Reports}, 4(1):4409, 2014.

\bibitem{andy2014}
Andy~M. Reynolds.
\newblock Distinguishing between lévy walks and strong alternative models:
  reply.
\newblock {\em Ecology}, 95(4):1109--1112, 2014.

\bibitem{simon2006}
Simon Benhamou.
\newblock Detecting an orientation component in animal paths when the preferred
  direction is individual-dependent.
\newblock {\em Ecology}, 87(2):518--528, 2006.

\bibitem{simon2007}
Simon Benhamou.
\newblock How many animals really do the lévy walk?
\newblock {\em Ecology}, 88(8):1962--1969, 2007.

\bibitem{zaburdaev2016}
V.~Zaburdaev, I.~Fouxon, S.~Denisov, and E.~Barkai.
\newblock Superdiffusive dispersals impart the geometry of underlying random
  walks.
\newblock {\em Phys. Rev. Lett.}, 117:270601, Dec 2016.

\end{thebibliography}
\end{document}